\RequirePackage{fix-cm} % Fix LaTeX2e bugs.
\documentclass[a4paper, twoside, reqno, dvips, 12pt]{amsart}
\usepackage{fixltx2e}     % Fix LaTeX2e bugs.

\usepackage[english]{babel}
\usepackage[latin1]{inputenc}

\usepackage{eucal}
\usepackage{esint}
\usepackage{xspace}
\usepackage{amsgen}
\usepackage{amssymb}
\usepackage{amsmath}
\usepackage{amsfonts}
\usepackage{MnSymbol}
\usepackage[nice]{nicefrac}
\usepackage{mathrsfs, dsfont}

\usepackage{a4wide}

\usepackage{indentfirst}
\usepackage{graphicx}
\usepackage{subfigure}
\usepackage{psfrag}

\usepackage{acronym}
\usepackage{microtype}
\usepackage[bf, up, hang]{caption}

\usepackage[usenames, dvipsnames, pdf]{pstricks}
\usepackage{epsfig}
\usepackage{pst-grad} % For gradients
\usepackage{pst-plot} % For axes

\usepackage{color}
\definecolor{oneblue}{rgb}{0.0, 0.0, 0.85}
\definecolor{darkgrey}{rgb}{0.273, 0.281, 0.30}

\usepackage[colorlinks,
            urlcolor=oneblue,
            linkcolor=oneblue,
            citecolor=oneblue,
            bookmarksopen=false,
            pagebackref]{hyperref}

\usepackage[compact]{titlesec}

\titleformat{\section}{\normalfont\Large\bfseries\sffamily\center\color{darkgrey}}{\thesection.}{0.5em}{}{}
\titleformat{\subsection}{\normalfont\large\bfseries\sffamily\color{darkgrey}}{\thesubsection.}{0.4em}{}{}
\titleformat{\subsubsection}{\normalfont\normalsize\bfseries\sffamily\color{darkgrey}}{\thesubsubsection.}{0.3em}{}{}

\titlespacing*{\section}{1.0em}{1.0em}{0.8em}[0em]
\titlespacing*{\subsection}{1.0em}{1.0em}{0.8em}[0em]
\titlespacing*{\subsubsection}{1.0em}{0.7em}{0.6em}[0em]

\usepackage{titletoc}

\contentsmargin{0.0em}
\dottedcontents{section}[2.5em]{\addvspace{0.5em}\bfseries}{1.0em}{0pc}
\dottedcontents{subsection}[4.5em]{}{2.6em}{0pc}
\dottedcontents{subsubsection}[5.5em]{}{3.0em}{0pc}

\usepackage{fancyhdr}
\usepackage{lastpage}

\newcommand*\Title{Computation of steady solitary gravity waves}
\newcommand*\Authors{D.~Dutykh \& D.~Clamond}

\numberwithin{equation}{section}

\newcommand{\sur}[1]{{#1}_\text{s}}                    % Surface.
\renewcommand{\bot}[1]{{#1}_\text{b}}                  % Bottom.

\newcommand{\depth}{d}
\newcommand{\s}{{\sf s}}
\newcommand{\ud}{\mathrm{d}}
\newcommand{\ui}{\mathrm{i}}
\newcommand{\ue}{\mathrm{e}}
\newcommand{\Ll}{\mathscr{L}}
\newcommand{\Nn}{\mathscr{N}}

\newcommand{\eps}{\varepsilon}
\renewcommand{\O}{\mathcal{O}}
\renewcommand{\L}{\mathcal{L}}

\renewcommand{\Re}{\operatorname{Re}}
\renewcommand{\Im}{\operatorname{Im}}

\newcommand{\half}{{\textstyle{1\over2}}}
\newcommand{\third}{{\textstyle{1\over3}}}

\begin{document}

\title[\Title]{Efficient computation of steady solitary gravity waves}

\author[D. Dutykh]{Denys Dutykh$^*$}
\address{University College Dublin, School of Mathematical Sciences, Belfield, Dublin 4, Ireland \and LAMA, UMR 5127 CNRS, Universit\'e de Savoie, Campus Scientifique, 73376 Le Bourget-du-Lac Cedex, France}
\email{Denys.Dutykh@univ-savoie.fr}
\urladdr{http://www.lama.univ-savoie.fr/~dutykh/}
\thanks{$^*$ Corresponding author}

\author[D. Clamond]{Didier Clamond}
\address{Laboratoire J.-A. Dieudonn\'e, Universit\'e de Nice -- Sophia Antipolis, Parc Valrose, 06108 Nice cedex 2, France}
\email{diderc@unice.fr}
\urladdr{http://math.unice.fr/~didierc/}

\maketitle
\thispagestyle{empty}

\begin{abstract}
An efficient numerical method to compute solitary wave solutions to the free surface Euler equations is reported. It is based on the conformal mapping technique combined with an efficient Fourier pseudo-spectral method. The resulting nonlinear equation is solved via the Petviashvili iterative scheme. The computational results are compared to some existing approaches, such as Tanaka's method and Fenton's high-order asymptotic expansion. Several important integral quantities are computed for a large range of amplitudes. The integral representation of the velocity and acceleration fields in the bulk of the fluid is also provided.

\bigskip
\noindent \textbf{\keywordsname:} Surface waves; gravity waves; Euler equations; solitary wave; Petviashvili method
\end{abstract}

\tableofcontents

\section{Introduction}

Solitary waves play a central role in nonlinear sciences \cite{Osborne2010}. They appear in various 
fields ranging from plasmas physics \cite{Lonngren1983} to hydrodynamics \cite{Kuznetsov1986} and 
nonlinear optics \cite{Gedalin1997}. For integrable models, it can be rigorously shown that any 
smooth and localised initial condition will split into a finite number of solitons plus a radiation 
\cite{Zabusky1965}. Solitons are special solitary waves interacting elastically, i.e., subject only 
to phase shifts after collisions \cite{John, Zabusky1965}. However, in full Euler equations, the 
interaction is known to be inelastic \cite{Seabra-Santos1989}. In some sense, solitons are elementary 
structures which span the system dynamics \cite{Miura1976} along with (relative) equilibria 
\cite{Patrick1992}, periodic orbits \cite{Cvitanovic1991}, etc. This is one of the main reasons why 
these solutions attract so much attention.

In some special cases the solitary waves can be found analytically. For example, explicit expressions 
are known for integrable models such as KdV and NLS equations \cite{Gardner1967, Gardner1974, Zakharov1974}, but also for some non-integrable Boussinesq-type \cite{Chen1998,C1,DMII} and Serre--Green--Naghdi \cite{Dias2010, Dutykh2011a, Green1976, Serre1953} equations. The examples of such analytical solutions are numerous \cite{Malfliet1992}. However, no closed-form solutions are known for the practically very important case of the free surface Euler equations. \textsc{Craig} \& \textsc{Sternberg} (1988) showed that solitary wave solutions to the Euler equations are necessarily positive and symmetric \cite{Craig1988} (without surface tension effects). In order to construct these solutions, one has to apply some approximate methods. Historically, high-order asymptotic approximations have been proposed first \cite{Fenton1972, Longuet-Higgins1974}. However, these solutions are asymptotic by construction and are therefore valid only in the limit $a/d \to 0$ ($a$ being the wave amplitude, $d$ the uniform undisturbed water depth); moreover, these series are known to be divergent \cite{Germain1967}. In order to avoid this limitation, several numerical approaches have been proposed \cite{Fenton1982,Okamoto2001} such as the Dirichlet-to-Neumann operator method \cite{Craig2002} or Boundary Integral Equation method \cite{Vanden-Broeck2007}. High 
amplitude solitary waves up to the limiting wave were studied by \textsc{Longuet-Higgins} \& \textsc{Tanaka} \cite{Longuet-Higgins1997}, among others. One of the most widely used methods nowadays is the Tanaka algorithm \cite{Tanaka1986}. In the present study, we are going to compare extensively our computational results to Tanaka's method.

The approach we proposed in a short recent preliminary study \cite{Clamond2012b} is also based on 
the conformal mapping technique, as the Tanaka method \cite{Tanaka1986}, for example. However, 
traditionally the conformal map is coupled with the Newton method \cite{Isaacson1966} to find the 
solitary wave profil \cite{Choi1999, Li2004, Milewski2010}. Newton-type iterations require the 
computation of a Jacobian matrix and the resolution of linear systems of equations (by direct or 
iterative methods) \cite{Trefethen1997}. From a computational point of view, simple iterative schemes 
are much easier to implement and they require only the evaluation of operators involved in the 
equation to be solved. In the previous study \cite{Clamond2012b}, we adopted the classical 
Petviashvili's iteration \cite{Petviashvili1976} in which the convergence is ensured by computing 
the so-called stabilising factor \cite{Pelinovsky2004}. This iterative scheme have been already 
applied to compute special solutions to many nonlinear wave equations \cite{Lakoba2007,Yang2010,
Fedele2011,Fedele2012}. An interesting comparison among different methods was recently performed 
for the solitary waves to the Benjamin equation \cite{Dougalis2012}. The combination of two main 
ingredients, i.e., the Petviashvili scheme together with the conformal mapping technique, allowed 
us to propose a very efficient numerical scheme for the computation of solitary gravity waves of 
the full Euler equations in the water of finite depth \cite{Clamond2012b}. The proposed algorithm 
admits a very compact and elegant implementation in \textsc{Matlab}, for example. The resulting 
script is ready to use and it can be freely downloaded from the \emph{Matlab Central} server 
\cite{Clamond2012}.

In the present study we perform further tests and validations of the new algorithm. Moreover, several 
important integral characteristics such as the mass, momentum, energy, \emph{etc}. are derived in the 
conformal space and computed numerically to the high accuracy for a wide range of solitary waves. Our 
method allows also to compute efficiently important physical fields in any point inside the bulk of 
the fluid layer. In this way, the pressure, velocities and accelerations are shown under a large 
amplitude solitary wave, up to an arbitrarily high accuracy.

This study is organised as follows. In Section~\ref{sec:model} we present the governing equations 
along with the conformal map technique. Several important integral quantities expressed in the 
transformed space are provided in Section~\ref{sec:intq}. The Babenko integral equation is derived 
in Section~\ref{sec:bab}. The following Section~\ref{sec:num} contains the description of the 
numerical scheme along with some validations and tests. Finally, some conclusions of this study 
are outlined in Section~\ref{sec:concl}.

\section{Mathematical model}\label{sec:model}

\begin{figure}
\centering
\scalebox{1} % Change this value to rescale the drawing.
{
\begin{pspicture}(0,-4.205)(9.541875,4.205)
\definecolor{color30}{rgb}{0.1803921568627451,0.18823529411764706,0.8941176470588236}
\definecolor{color136g}{rgb}{0.3333333333333333,0.17647058823529413,0.07058823529411765}
\definecolor{color136f}{rgb}{0.8196078431372549,0.6196078431372549,0.49019607843137253}
\definecolor{color136}{rgb}{0.3176470588235294,0.1568627450980392,0.08235294117647059}
\definecolor{color426}{rgb}{0.00392156862745098,0.00392156862745098,0.00392156862745098}
\psline[linewidth=0.03cm,linestyle=dashed,dash=0.16cm 0.16cm,arrowsize=0.05291667cm 2.0,arrowlength=1.4,arrowinset=0.4]{->}(0.0,2.95)(8.36,2.95)
\psline[linewidth=0.03cm,linestyle=dashed,dash=0.16cm 0.16cm,arrowsize=0.05291667cm 2.0,arrowlength=1.4,arrowinset=0.4]{<-}(4.26,4.19)(4.24,0.63)
\psline[linewidth=0.04cm,linecolor=color30](0.38,2.95)(2.62,2.95)
\psline[linewidth=0.04cm,linecolor=color30](5.84,2.95)(8.06,2.95)
\pscustom[linewidth=0.04,linecolor=color30]
{
\newpath
\moveto(2.5943189,2.9588428)
\lineto(2.8336933,3.0063117)
\curveto(2.95338,3.030046)(3.1584277,3.103938)(3.2437882,3.1540952)
\curveto(3.3291485,3.2042522)(3.5201929,3.3300357)(3.6258764,3.4056618)
\curveto(3.73156,3.481288)(3.966687,3.5701487)(4.0961304,3.5833836)
\curveto(4.2255735,3.5966182)(4.4785905,3.571543)(4.6021643,3.533233)
\curveto(4.7257385,3.4949229)(4.9376583,3.393094)(5.026004,3.3295753)
\curveto(5.11435,3.2660565)(5.2962804,3.1544218)(5.3898644,3.1063058)
\curveto(5.483449,3.0581896)(5.6313233,2.9884672)(5.794192,2.9236476)
}
\psframe[linewidth=0.02,linecolor=color136,dimen=outer,fillstyle=gradient,gradlines=2000,gradbegin=color136g,gradend=color136f,gradmidpoint=1.0](7.96,1.11)(0.36,0.87)
\usefont{T1}{ptm}{m}{n}
\rput(8.181406,2.735){$x$}
\usefont{T1}{ptm}{m}{n}
\rput(3.9114063,3.995){$y$}
\usefont{T1}{ptm}{m}{n}
\rput(4.0314064,2.695){$O$}
\usefont{T1}{ptm}{m}{n}
\rput(6.9414062,1.375){$y = -d$}
\usefont{T1}{ptm}{m}{n}
\rput(7.0514064,3.195){$y = \eta(x)$}
\usefont{T1}{ptm}{m}{n}
\rput(1.8214062,1.915){$u\to -c\ \text{as}\ x\to\pm\infty$}
\psline[linewidth=0.03cm,linestyle=dashed,dash=0.16cm 0.16cm,arrowsize=0.05291667cm 2.0,arrowlength=1.4,arrowinset=0.4]{<-}(4.24,-1.29)(4.26,-4.19)
\usefont{T1}{ptm}{m}{n}
\rput(3.9114063,-2.385){$O$}
\psline[linewidth=0.03cm,linestyle=dashed,dash=0.16cm 0.16cm,arrowsize=0.05291667cm 2.0,arrowlength=1.4,arrowinset=0.4]{->}(0.52,-2.15)(8.4,-2.13)
\usefont{T1}{ptm}{m}{n}
\rput(8.351406,-2.385){$\alpha$}
\usefont{T1}{ptm}{m}{n}
\rput(3.9314063,-1.465){$\beta$}
\psframe[linewidth=0.02,linecolor=color136,dimen=outer,fillstyle=gradient,gradlines=2000,gradbegin=color136g,gradend=color136f,gradmidpoint=1.0](8.08,-3.75)(0.48,-3.99)
\psline[linewidth=0.04cm,linecolor=color30](0.8,-2.15)(7.9,-2.13)
\usefont{T1}{ptm}{m}{n}
\rput(6.4414062,-1.905){$\beta = 0$}
\usefont{T1}{ptm}{m}{n}
\rput(6.521406,-3.525){$\beta = -d$}
\pscustom[linewidth=0.04,linecolor=color426]
{
\newpath
\moveto(5.08,0.17)
\lineto(5.159655,0.08058838)
\curveto(5.1994824,0.03588257)(5.250172,-0.07588257)(5.2610345,-0.14294128)
\curveto(5.271897,-0.21)(5.2863793,-0.34411743)(5.29,-0.41117644)
\curveto(5.2936206,-0.47823548)(5.29,-0.6123529)(5.2827587,-0.67941165)
\curveto(5.2755175,-0.7464709)(5.250172,-0.8526477)(5.1958623,-0.97)
}
\psline[linewidth=0.04cm,linecolor=color426,arrowsize=0.05291667cm 2.0,arrowlength=1.4,arrowinset=0.4]{->}(5.2,-0.95)(5.02,-1.23)
\psline[linewidth=0.04cm,linecolor=color426,arrowsize=0.05291667cm 2.0,arrowlength=1.4,arrowinset=0.4]{->}(5.1,0.15)(4.9,0.39)
\usefont{T1}{ptm}{m}{n}
\rput(7.2814064,-0.265){$x = \alpha + X(\alpha,\beta)$}
\usefont{T1}{ptm}{m}{n}
\rput(7.251406,-0.725){$y = \beta + Y(\alpha,\beta)$}
\end{pspicture}
}
\caption{\em Definition sketch of the physical and transformed domains.}
\label{fig:sketch}
\end{figure}
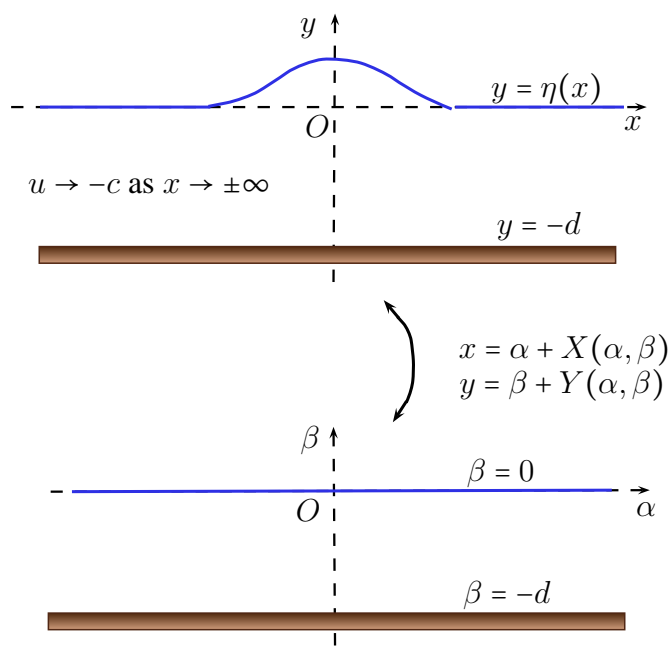

We consider steady two-dimensional potential flows due to surface gravity solitary waves in 
constant depth. The fluid is homogeneous, the pressure is zero at the impermeable free surface 
and the seabed is fixed, horizontal and impermeable.

Let be $(x,y)$ a Cartesian coordinate system moving with the wave, $x$ being the horizontal 
coordinate and $y$ the upward vertical one. Since solitary waves are localised in space, the 
surface elevation tends to zero, along with all derivatives, as $x \to \pm\infty$, and $x=0$ 
is the abscissa of the crest. The equations of the bottom, of the free surface and of the mean 
water level are given correspondingly by $y=-\depth$, $y=\eta(x)$ and $y=0$. The parameter 
$a\equiv\eta(0)$ denotes the wave amplitude. Since gravity solitary waves of the Euler equations 
are known to be symmetric and positive \cite{Craig1988}, we have $\eta(-x) = \eta(x) \geqslant 0$ 
and $a = \max(\eta)$.

Let be $\phi$, $\psi$, $u$ and $v$ the velocity potential, the stream function, the horizontal 
and vertical velocities, respectively, such that $u=\phi_x=\psi_y$ and $v=\phi_y=-\psi_x$. It 
is convenient to introduce the complex potential $f\equiv\phi+\ui\/\psi$ (with $\ui^2=-1$) and 
the complex velocity $w\equiv u-\ui\/v$ that are holomorphic functions of $z\equiv x+\ui\/y$ 
(i.e., $w = \ud f/\ud z$). The complex conjugate is denoted with a star (e.g., $z^\ast=x-\ui\/y$), 
while subscripts `b' denote the quantities written at the seabed --- e.g., $\bot{z}(x) = x-\ui\depth$, 
$\bot{\phi}(x)=\phi(x,y\!=\!-\depth)$ --- and subscripts `s' denote the quantities written at 
the free surface --- e.g., $\sur{z}(x) = x + \ui\eta(x)$, $\sur{\phi}(x) = \phi(x,y\!=\!\eta(x))$. 
Note that, e.g., $\sur{u} = \sur{(\partial_x\phi)} \neq \partial_x(\sur{\phi}) = \sur{u} + \eta_x\sur{v}$. 
We also emphasise that $\sur{\psi}$ and $\bot{\psi}$ are constants because the surface and the bottom 
are streamlines.

The far field velocity is such that  $(u,v) \to (-c,0)$ as $x \to \pm\infty$, so $c$ is the wave 
phase velocity observed in the frame of reference where the fluid is at rest at infinity ($c>0$ 
if the wave travels to the increasing $x$-direction). Note that $c = (\bot{\psi} - \sur{\psi})/\depth$ 
due to the mass conservation.

The dynamic condition can be expressed in form of the Bernoulli equation
\begin{equation}\label{eq:bernbase}
2\,p\ +\ 2\,g\,y\ +\ u^2\ +\ v^2\ =\ c^2,
\end{equation}
where $p$ is the pressure divided by the constant density $\rho$ and $g > 0$ is the acceleration 
due to gravity. At the free surface the pressure equals that of the atmosphere which is constant 
and set to zero without loss of generality, i.e., $\sur{p} = 0$.

\subsection{Conformal mapping}\label{sec:conmap}

Let be the change of independent variable $z\mapsto\zeta\equiv(\ui\sur{\psi}-f)/c$, that conformally 
maps the fluid domain $\{-\infty\leqslant x\leqslant\infty; -\depth\leqslant y\leqslant\eta\}$ into 
the strip $\{-\infty \leqslant \alpha \leqslant\infty; -\depth\leqslant\beta\leqslant 0\}$ where 
$\alpha\equiv \Re(\zeta)$ and $\beta\equiv\Im(\zeta)$ (c.f. Figure~\ref{fig:sketch}). The conformal 
mapping yields the Cauchy--Riemann relations $x_\alpha = y_\beta$ and $x_\beta = -y_\alpha$, while 
the complex velocity along with its components are
\begin{equation*}
\frac{w}{c}\ =\ -\left(\frac{\ud\,z}{\ud\/\zeta}\right)^{\!-1}, \qquad
\frac{u}{c}\ =\ \frac{-\,x_\alpha}{x_\alpha^{\,2}\/+\/y_\alpha^{\,2}}, \qquad
\frac{v}{c}\ =\ \frac{-\,y_\alpha}{x_\alpha^{\,2}\/+\/y_\alpha^{\,2}}, \qquad
\frac{u^2\,+\,v^2}{c^2}\ =\ \frac{1}{x_\alpha^{\,2}\/+\/y_\alpha^{\,2}}. 
\end{equation*}

Let us introduce new dependent variables $X(\alpha,\beta)\equiv x-\alpha$ and $Y(\alpha, \beta) 
\equiv y-\beta$, so the Cauchy--Riemann relations $X_\alpha=Y_\beta$ and $X_\beta=-Y_\alpha$ hold, 
while the bottom ($\beta = -\depth$) and the free surface ($\beta=0$) impermeabilities yield 
$\bot{Y}(\alpha)\equiv Y(\alpha,-\depth) = 0$ and $\sur{Y}(\alpha)\equiv Y(\alpha,0) = \eta$. At 
the crest and in the far field (i.e., $\alpha=0$ and $\alpha=\pm\infty$), we have from the 
$X(0,\beta) = 0$ and $X(\pm\infty,\beta) = \pm X_{\infty}$, thence $X$ is a bounded odd function.

The functions $X$ and $Y$ can be expressed in term of $\bot{X}$ solely \cite{Clamond1999, Clamond2003}
\begin{align}
X(\alpha,\beta)\ =&\ \Re\left\{\bot{X}(\zeta+\ui\depth)\right\}\ =\ 
\cos\!\left[\,(\beta + \depth)\/\partial_\alpha\,\right]\bot{X}(\alpha), \label{eq:solxxbot}\\
Y(\alpha,\beta)\ =&\ \Im\left\{\bot{X}(\zeta+\ui\depth)\right\}\ =\ 
\sin\!\left[\,(\beta+\depth)\/\partial_\alpha\,\right]\bot{X}(\alpha). \label{eq:solyxbot}
\end{align}
Thus, the Cauchy--Riemann relations and the bottom impermeability are fulfilled identically. At the 
free surface $\beta=0$, \eqref{eq:solxxbot} yields $\sur{X}(\alpha)=\cos\!\left[\depth\partial_\alpha 
\right]\bot{X}(\alpha)$, that can be inverted as $\bot{X}(\alpha)= \sec\!\left[\depth\partial_\alpha 
\right] \sur{X}(\alpha)$, and hence the relation \eqref{eq:solyxbot} yields
\begin{equation}\label{eq:relYXs}
\sur{Y}(\alpha)\ =\ \tan\!\left[\,\depth\,\partial_\alpha\,\right]\sur{X}(\alpha),
\end{equation}
which relates quantities written at the free surface only. The relation \eqref{eq:relYXs} can be 
trivially inverted giving, in particular, $\sur{(\partial_\alpha X)}=\mathscr{C}\{\sur{Y}\} \equiv 
\partial_\alpha\cot\!\left[\/\depth\/\partial_\alpha\/\right]\sur{Y}$, where $\mathscr{C}$ is a 
self-adjoint positive-definite pseudo-differential operator acting on a pure frequency as
\begin{equation}\label{eq:cop}
\mathscr{C}\left\{\ue^{\ui k\alpha}\right\}\,=\,\left\{
\begin{array}{lr}
k\coth(k\depth)\,
\ue^{\ui k\alpha} &\quad (k\neq0)  \\
1/\depth & \quad  (k=0)
\end{array}
\right.
\end{equation}
This operator can be efficiently evaluated in the Fourier space using a FFT algorithm.

\subsection{Integral quantities}\label{sec:intq}

The wave can be characterised by several integral parameters \cite{LH, McCowan1891, Starr1947}. 
These quantities are defined in the frame of reference where the flow is at rest as $x\to\pm\infty$. 
This choice is made since the kinetic energy, for example, is infinite in the reference frame moving 
with a solitary wave. The main integral quantities of interest here are
\begin{align*}
\renewcommand{\arraystretch}{1.7}
\begin{array}{rlc}
\textit{\small Wave Mass:}& \mathcal{M}\ \equiv\ \int_{-\infty}^{\infty}\eta\,\ud\/x\ 
=\ \int_{-\infty}^{\infty}(1+\mathscr{C}\{\eta\})\,\eta\,\ud\/\alpha, & (\mbox{\sc i.m}) \\
\textit{\small Circulation:}& \mathcal{C}\ \equiv\ \int_{-\infty}^{\infty}(\sur{u}+c+\sur{v}\eta_x)\,\ud\/x\ 
=\ \int_{-\infty}^{\infty}c\,\mathscr{C}\{\eta\}\,\ud\/\alpha\ 
=\ c\left[\/\sur{X}\/\right]_{-\infty}^{+\infty}, &(\mbox{\sc i.c})\\
\textit{\small Impulse:}& \mathcal{I}\ \equiv\ \int_{-\infty}^{\infty}\int_{-\depth}^\eta\,
(u+c)\,\ud\/y\,\ud\/x\ =\ c\,\mathcal{M}\ =\ \depth\,\mathcal{C}\ +\ \int_{-\infty}^{\infty}
c\,\eta\,\mathscr{C}\{\eta\}\,\ud\/\alpha, &(\mbox{\sc i.i})\\
%
%\text{\small Angular Momentum:}& \mathcal{A}\ \equiv\ \int_{-\infty}^{\infty}\int_{-\depth}^\eta\, 
%[\/y\/(u+c)\/-\,x\/v\/]\,\ud\/y\,\ud\/x\ =\ \\
%
\textit{\small Kinetic Energy:}& \mathcal{K}\ \equiv\ \int_{-\infty}^{\infty}\int_{-\depth}^{\eta}
\half\,[\,(u+c)^2+v^2\,]\,\ud\/y\,\ud\/x\ =\ \half\,c\,(\,\mathcal{I}\,-\,\depth\,\mathcal{C}\,),  
&(\mbox{\sc i.k})\\ 
\textit{\small Potential Energy:}& \mathcal{V}\ \equiv\ \int_{-\infty}^{\infty}\half\,g\,\eta^2\,\ud\/x\ 
=\ \int_{-\infty}^{\infty}\half\,g\,\eta^2\,(1+\mathscr{C}\{\eta\})\,\ud\/\alpha
\ =\ \third\,(\/c^2\/-\/g\/\depth\/)\,\mathcal{M}, & (\mbox{\sc i.v})\\
\textit{\small Total Energy:}& \mathcal{E}\ \equiv\ \mathcal{K}\ +\ \mathcal{V},  &(\mbox{\sc i.e})\\
\textit{\small Energy Flux:}& \mathcal{F}\ \equiv\ \int_{-\infty}^{\infty}\int_{-\depth}^\eta \left[\/p\/
+\/g\/y\/+\/\half\/(u+c)^2\/+\/\half\/v^2\/\right](u+c)\, \ud\/y\,\ud\/x\ =\ c\,\mathcal{E}, 
&(\mbox{\sc i.f}) \\
\textit{\small Group Velocity:}& c_\text{g}\ \equiv\ \mathcal{F}\,/\,\mathcal{E}\ =\ c, 
&(\mbox{\sc i.g})\\
\textit{\small Lagrangian:}& \L\ \equiv\ \mathcal{K}\ -\ \mathcal{V}\ =\ 
\int_{-\infty}^{\infty}\,\mathfrak{L}\ \ud\/\alpha, &(\mbox{\sc i.l})
\end{array}
\renewcommand{\arraystretch}{1.0}
\end{align*}
$\mathfrak{L}$ being the Lagrangian density \cite{Okamoto2001} defined from the integral relations above, i.e.,
\begin{equation}\label{lagden}
\mathfrak{L}\ =\ \half\,c^2\,\eta\,\mathscr{C}\{\eta\}\ -\ \half\,g\,\eta^2\,
(\/1\,+\,\mathscr{C}\{\eta\}\/).
\end{equation}
The equalities in the integral relations above are easily obtained via some trivial derivations \cite{LH, McCowan1891, Starr1947}.

Luke's Lagrangian for water waves \cite{Clamond2009, Cotter2010, Luke1967} reduces to the Hamilton principle --- i.e., the kinetic minus potential energies --- if the Laplace equation together with the bottom and surface impermeabilities are identically fulfilled. This is precisely the case when using the conformal mapping and the relations derived in Section~\ref{sec:conmap}, leading in particular to the relation $\mathcal{K} = \int_{-\infty}^{\infty} \half\,c^2\,\eta\,\mathscr{C}\{\eta\}\,\ud\/\alpha$ which can then be substituted into the Lagrangian $\L$. Conversely, the last relation in ({\sc i.v}) holds only if the equation for the momentum flux is fulfilled. Thus, the last relation in ({\sc i.v}) cannot be substituted into $\L$, but it can be used to monitor the accuracy of any 
resolution procedure.

\subsection{Babenko's equation}\label{sec:bab}

Since we have a Lagrangian at our disposal, an equation for $\eta$ can be obtained from the variational principle $\delta\mathcal{L}=0$ leading to the following Euler--Lagrange equation
\begin{equation}\label{eqbab}
0\ =\ \frac{\partial\,\mathfrak{L}}{\partial\/\eta}\ +\ \mathscr{C}\!\left\{\frac{\partial\,
\mathfrak{L}}{\partial\,\mathscr{C}\{\eta\}}\right\}\ =\ c^2\,\mathscr{C}\{\eta\}\ -\ g\,\eta\ 
-\ \half\, g\,\mathscr{C}\{\eta^2\}\ -\ g\,\eta\,\mathscr{C}\{\eta\},
\end{equation}
which is the Babenko equation for gravity solitary surface waves \cite{Babenko1987}. By applying the operator $\mathscr{C}^{-1}$ to the previous equation and  splitting the linear and nonlinear parts, one obtains an equivalent version which is more convenient for numerical computations
\begin{align}\label{eq:babpet}
c^2\,\eta\ -\ g\,\mathscr{C}^{-1}\{\eta\}\ =\ \half\,g\,\eta^2\ +\ 
g\,\mathscr{C}^{-1}\{\eta\,\mathscr{C}\{\eta\}\}.
\end{align}

The numerical resolution of \eqref{eq:babpet} is explained in the Section~\ref{sec:num} below.

\subsection{Velocity and pressure fields in the fluid}\label{sec:bulk}

In the numerical procedure described below, we use conformal mapping and a Fourier pseudo-spectral method to solve the equations. This means that we obtain a discrete approximation equally spaced along each streamline. However, for practical applications, it is often necessary to determine the fields (velocity, pressure, etc.) at various positions that are not necessarily the nodes used for the computation. These informations can be obtained as follows.

Let be $W(z) = c + w(z)$ the complex velocity observed in the frame of reference where the fluid is at rest in the far field (i.e., $W \to 0$ as $\Re z \to \pm\infty$). The complex velocity being known at the fluid boundaries from our approximation procedure, $W$ at any complex abscissa $z$ can be obtained from the Cauchy integral
\begin{equation}\label{cauchint}
\ui\,\theta\,W(z)\ =\ \mathrm{P.V.}\ointctrclockwise\frac{c+w(z_1)}{z_1-z}\,\ud\/z_1,
\end{equation}
where $\theta=2\pi$ if $z$ is strictly inside the fluid domain (i.e., $\Im(z)<\eta$), $\theta=\pi$ 
if $z$ is at the free surface (i.e., $\Im(z)=\eta$) and $\theta=0$ if $z$ is strictly above the free surface (i.e., $\Im(z)>\eta$). The bottom impermeability being taken into account via the method of images (Schwartz reflection principle \cite{Lavrent'ev1967}), the Cauchy integral \eqref{cauchint} yields for any $z$ below the surface
\begin{align*}
W(z)\ =\ \frac{\ui\/c}{2\/\pi}\int_{-\infty}^{\infty}\left[\,\frac{\sur{z}'(\alpha)\,-\,1}
{\sur{z}(\alpha)\,-\,z}\ -\ \frac{\sur{z}'^\ast(\alpha)\,-\,1}{\sur{z}^\ast(\alpha)\,
-\,2\/\ui\/\depth\,-\,z}\,\right]\ud\/\alpha,
\end{align*}
where $\sur{z}(\alpha) = \alpha+\sur{X}(\alpha)+\ui\eta(\alpha)$ and $\sur{z}'(\alpha) = \ud\sur{z} /\ud\alpha = 1 + \mathscr{C}\{\eta\}(\alpha) + \ui\eta_\alpha(\alpha)$, $\sur{X}$ and $\eta$ being known from the numerical resolution of the Babenko equation. From this relation, we obtain the derivative of $W$ (required to compute the acceleration field)
\begin{align*}
\frac{\ud\,W(z)}{\ud\/z}\ =\ \frac{\ui\/c}{2\/\pi}\int_{-\infty}^{\infty}\left[\,
\frac{\sur{z}'(\alpha)\,-\,1}{(\,\sur{z}(\alpha)\,-\,z\,)^2}\ -\ \frac{\sur{z}'^\ast(\alpha)\,-\,1}
{(\,\sur{z}^\ast(\alpha)\,-\,2\/\ui\/\depth\,-\,z\,)^2}\,\right]\ud\/\alpha,
\end{align*}
and the complex potential
\begin{align*}
\Xi(z)\ =\ \int_{-\infty}^{\infty}\left\{\,\frac{\sur{z}'(\alpha)-1}{2\/\pi\,/\,\ui\/c}
\log\left(\frac{\sur{z}(\alpha)+\ui\depth}{\sur{z}(\alpha)-z}\right)\, +\ 
\left[\,\frac{\sur{z}'(\alpha)-1}{2\/\pi\,/\,\ui\/c}\log\left(\frac{\sur{z}(\alpha)+\ui\depth}
{\sur{z}(\alpha)+2\ui\depth-z^\ast}\right)\,\right]^\ast\,\right\}\ud\/\alpha,
\end{align*}
such that $W = \ud\Xi/\ud z$ and $\Im(\Xi) = 0$ at the bed. Below, we will present some numerical results using these analytical representations.

\section{Numerical scheme}\label{sec:num}

The Babenko equation \eqref{eq:babpet} has a major advantage with respect to the original 
Bernoulli integral \eqref{eq:bernbase} --- this equation is posed on the fixed domain in 
conformal variables and is only quadratic in nonlinearities. For the sake of convenience 
we separate the linear and nonlinear parts of equation \eqref{eq:babpet} as
\begin{equation}\label{eq:pet}
\Ll\{\eta\}\, =\, \Nn\{\eta\}, \qquad
\Ll\{\eta\}\, \equiv\, c^2\,\eta\, -\, g\,\mathscr{C}^{-1}\{\eta\}, \quad
\Nn\{\eta\}\, \equiv\, g\,\mathscr{C}^{-1}\!\left\{\,\eta\,\mathscr{C}\{\eta\}
\,\right\}\, +\, \half\,g\,\eta^2.\ 
\end{equation}
This is the equation we are solving numerically.

\subsection{Petviashvili's iterations}

In order to solve numerically the equation \eqref{eq:pet}, we apply the classical Petviashvili 
scheme \cite{Lakoba2007, Petviashvili1976, Yang2010}:
\begin{equation}\label{eq:pet2}
\eta_{n+1}\ =\ S_n^{\,2}\,\Ll^{-1}\circ\Nn\{\eta_n\}, \qquad
S_n\ =\ \frac{\int_{-\infty}^\infty\hat{\eta}_n^\ast\,\mathfrak{F}\{\Ll\{\eta_n\}\}\,
\ud\/k}{\int_{-\infty}^\infty\hat{\eta}_n^\ast\,\mathfrak{F}\{\Nn\{\eta_n\}\}\,
\ud\/k}\ =\ \frac{\int_{-\infty}^\infty{\eta}_n\,\Ll\{\eta_n\}\,
\ud\/\alpha}{\int_{-\infty}^\infty{\eta}_n\,\Nn\{\eta_n\}\,
\ud\/\alpha},
\end{equation}
where $S_n$ is the co-called {\em stabilisation factor\/} which can be computed in the real or Fourier space (the equality following from the Parseval identity \cite{Titchmarsh1976}), the Fourier transform being
\begin{equation*}
\hat{f}(k)\ =\ \mathfrak{F}\{f\}\ = 
\int_{-\infty}^\infty f(\alpha)\,\ue^{-\ui k\alpha}\ \ud\/\alpha, \qquad
f(\alpha)\ =\ \mathfrak{F}^{-1}\{\hat{f}\}\ =\ 
\frac{1}{2\pi}\int_{-\infty}^\infty\hat{f}(k)\,\ue^{\ui k\alpha}\ \ud\/k.
\end{equation*}
We will systematically privilege the Fourier space since the operators $\mathscr{C}$ and 
$\mathscr{C}^{-1}$ can be very efficiently computed according to definition \eqref{eq:cop} 
using the FFT algorithm \cite{Cooley1965}.

The convergence of the iterative process \eqref{eq:pet2} is checked by following the norm 
of the difference between two successive iterations along with the residual in the $\ell_\infty$ norm
\begin{equation*}
\|\,\eta_{n+1}\, -\, \eta_{n}\, \|_\infty\ <\ \eps_1, \qquad
\|\,\Ll\{\eta_n\}\, -\, \Nn\{\eta_n\}\,\|_\infty\ <\ \eps_2,
\end{equation*}
where $\eps_{1,2}$ are some prescribed tolerance parameters that are usually of the order 
of the floating point arithmetics precision. Pelinovsky and Stepanyants \cite{Pelinovsky2004} showed that the Petviashvili scheme has the linear rate of convergence. Some modifications of the classical Petviashvili scheme have been proposed in the literature \cite{Lakoba2007}. However, we found the convergence of the classical scheme completely satisfactory for gravity waves (see Section \ref{sec:res} for more details). So, at the current stage, no significant improvements are necessary.

\subsection{Initial guess}

There are several possibilities to choose the initial guess of the free surface elevation 
$\eta_{0}(\alpha)$. One of the simplest possibilities consists in taking a KdV-like analytical approximation
\begin{equation}\label{eq:serre}
\frac{\eta_0}{\depth}\ \approx\ \frac{F^2\,-\,q_\text{c}^{\,2}}{1\,+\,\cosh(\kappa\alpha)}, 
\qquad F^2\ =\ \frac{\tan(\kappa\depth)}{\kappa\depth}\ =\ q_\text{c}^{\,2}\,
+\,\frac{2\/a}{\depth}, \qquad (\kappa\/d)^2\ \approx\ \frac{3\,a}{d+a}, 
\end{equation}
where $F = c/\sqrt{g\depth}$ (with $|F|\geqslant1$) is a Froude number and $q_\text{c} = 
-u_\text{c}/\sqrt{gd}$ (with $0 \leqslant |q_\text{c}| \leqslant1$), $u_\text{c}$ being the 
fluid horizontal velocity at the wave crest. Values of $|q_\text{c}|$ close to $1$ correspond to infinitesimal waves, while $q_\text{c} = 0$ is the limiting case where the crest becomes a stagnation point with a $120^\circ$ inner angle. The double equality in \eqref{eq:serre} is exact, the first equality was derived by \textsc{McCowan} \cite{McCowan1891}, the second one being the Bernoulli equation written at the wave crest. A more accurate choice could be a high-order asymptotic approximation \cite{Fenton1972}. However, our numerical tests have shown a negligible sensitivity of the algorithm with respect to the initial guess --- with almost any reasonable choice the iterative scheme converged to the right solution with the same rate. Consequently, we always use the simplest analytical solution \eqref{eq:serre} to initialise the iterative process.

The relations \eqref{eq:serre} are convenient if the wave is defined by its dimensionless amplitude $a/\depth$. If this is not the case, one has to solve equations to find the parameters. In order to avoid this unnecessary overhead, one can use simple approximations for the initial guess. For instance, if the wave is defined by its Froude number $F$, one can use the approximations 
\begin{equation*}
a\,/\,\depth\ \approx\,\left(F^2-1\right)\left/\,\left(2-F^2\right),\right. \qquad
q_\text{c}^{\,2}\ \approx\ \left(2-F^4\right)\left/\,\left(2-F^2\right).\right.
\end{equation*}
Conversely, if the wave is defined by the parameter $q_\text{c}$ (more suitable for large waves), we have 
\begin{equation*}
4\,a\,/\,\depth\ \approx\ \sqrt{\,8\,-\,8\,q_\text{c}^{\,2}\,+\,q_\text{c}^{\,4}\,}\ -\ q_\text{c}^{\,2}.
\end{equation*}
However, in numerical computations presented below we use the simple parametrization of the travelling wave solution in terms of the Froude number $F$ and formulas \eqref{eq:serre} for the initial guess.

\subsection{Numerical results}\label{sec:res}

The validation of the proposed algorithm is done by comparing it with some existing approaches. One of the most used algorithms nowadays is the one of Tanaka \cite{Tanaka1986}, also based on the conformal mapping technique. Tanaka's solution is parametrized by the dimensionless parameter $q_\text{c}$ defined above. Let us take, for example, $q_\text{c}=0.87$, which corresponds to a mild solitary wave. Tanaka's algorithm was implemented in \textsc{Matlab}\footnote{In the sequel, all the algorithms will be compared in the same computing environment, i.e. \textsc{Matlab}.} in its original version without any peculiar optimisations. The interval $-30 \leqslant \phi \leqslant30$ 
is discretised using $20,001$ points and the iterations are continued until the tolerance $10^{-10}$ is achieved between two successive values of the Froude number:
\begin{equation*}
|\,F_{n+1}\,-\,F_n\,|\ <\ \eps_3\ \ (=\ 10^{-10} \ \text{in our computation}).
\end{equation*}
This computation required 173 iterations which lasted $1,269\,\textsf{s}$ on our desktop computer. The resulting Froude number computed by Tanaka's algorithm is $F_{\mbox{\tiny\sc Tan}} = 1.066365888477383$.

\begin{figure}
  \centering
  \subfigure[]%
  {\includegraphics[width=0.49\textwidth]{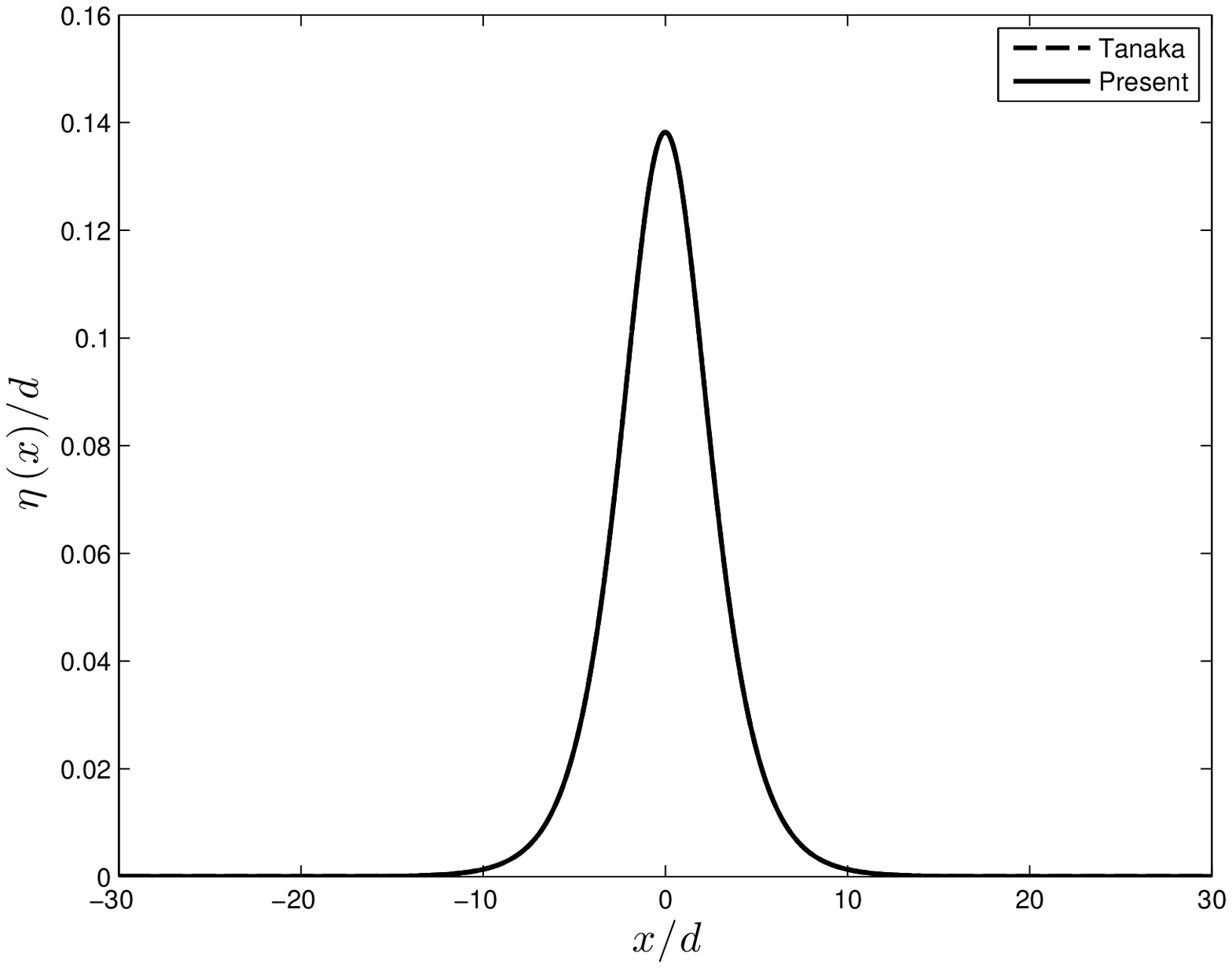}}
  \subfigure[]%
  {\includegraphics[width=0.49\textwidth]{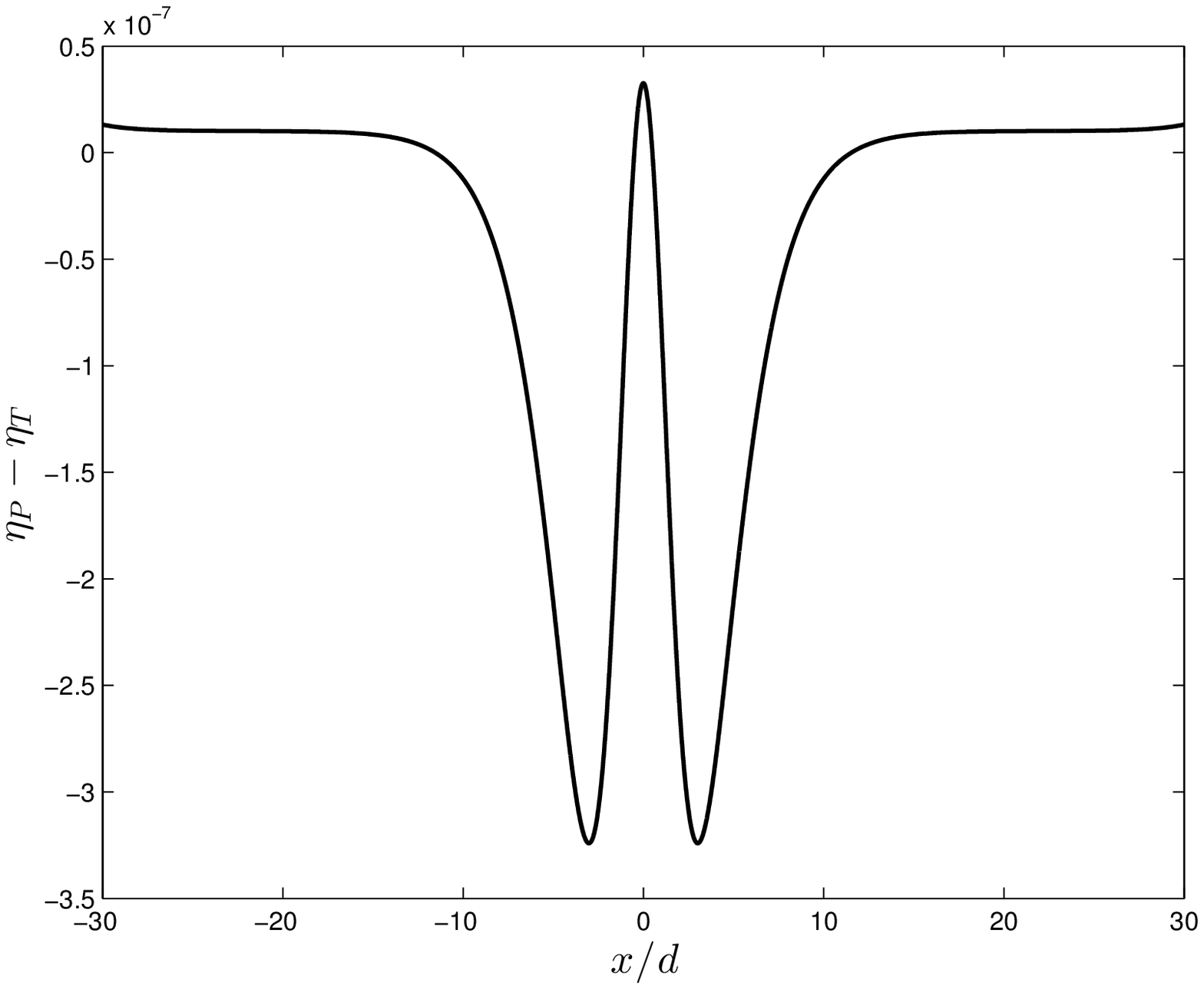}}
  \caption{\em\small Comparison between Tanaka's and the present solutions for the same Froude 
  number $F_T = 1.066365888477383$. Left: free surface; Right: difference (the vertical scale 
  is $10^{-7}$).} \label{fig:small}
\end{figure}

\begin{figure}
  \centering
  \subfigure[]%
  {\includegraphics[width=0.49\textwidth]{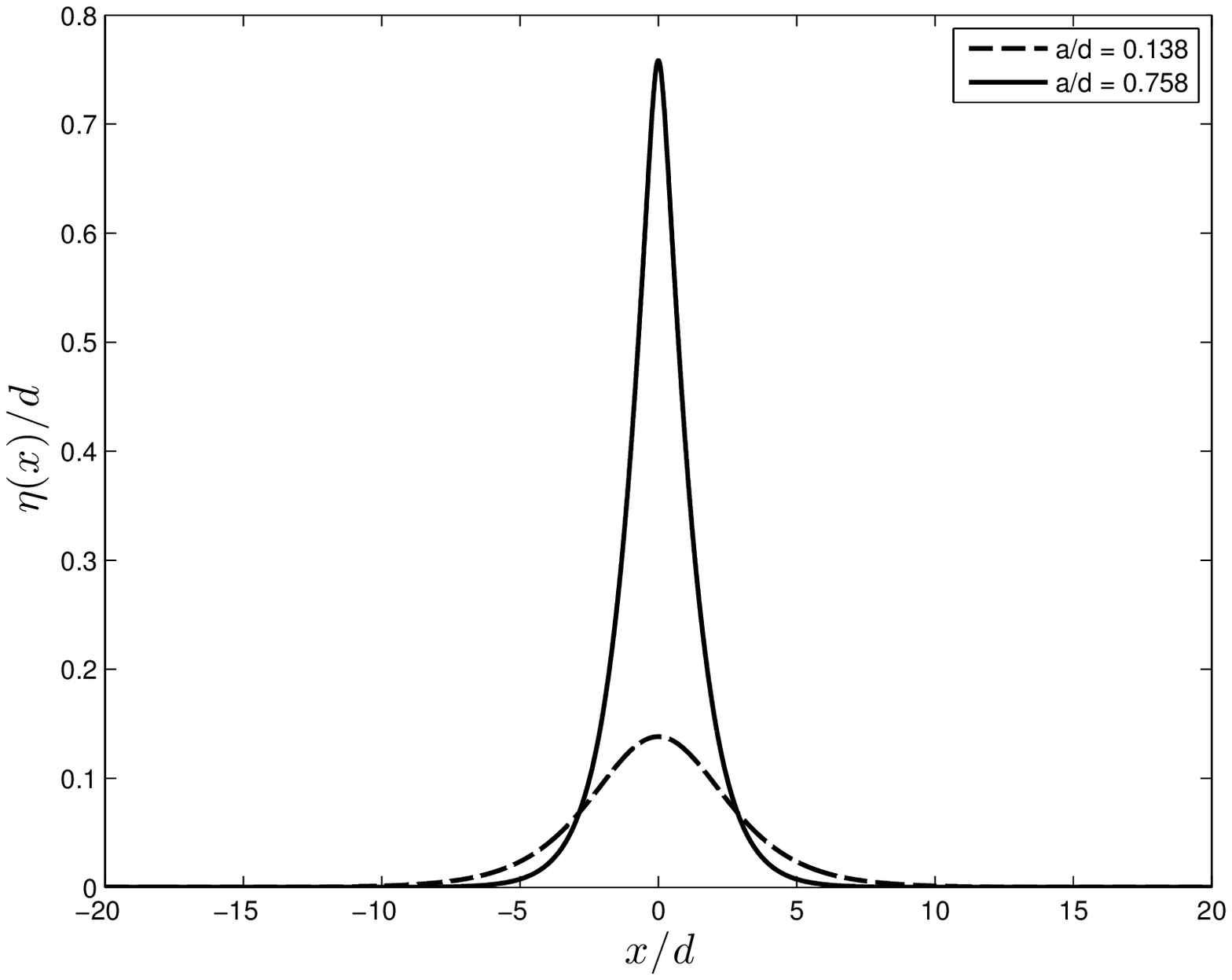}}
  \subfigure[]%
  {\includegraphics[width=0.49\textwidth]{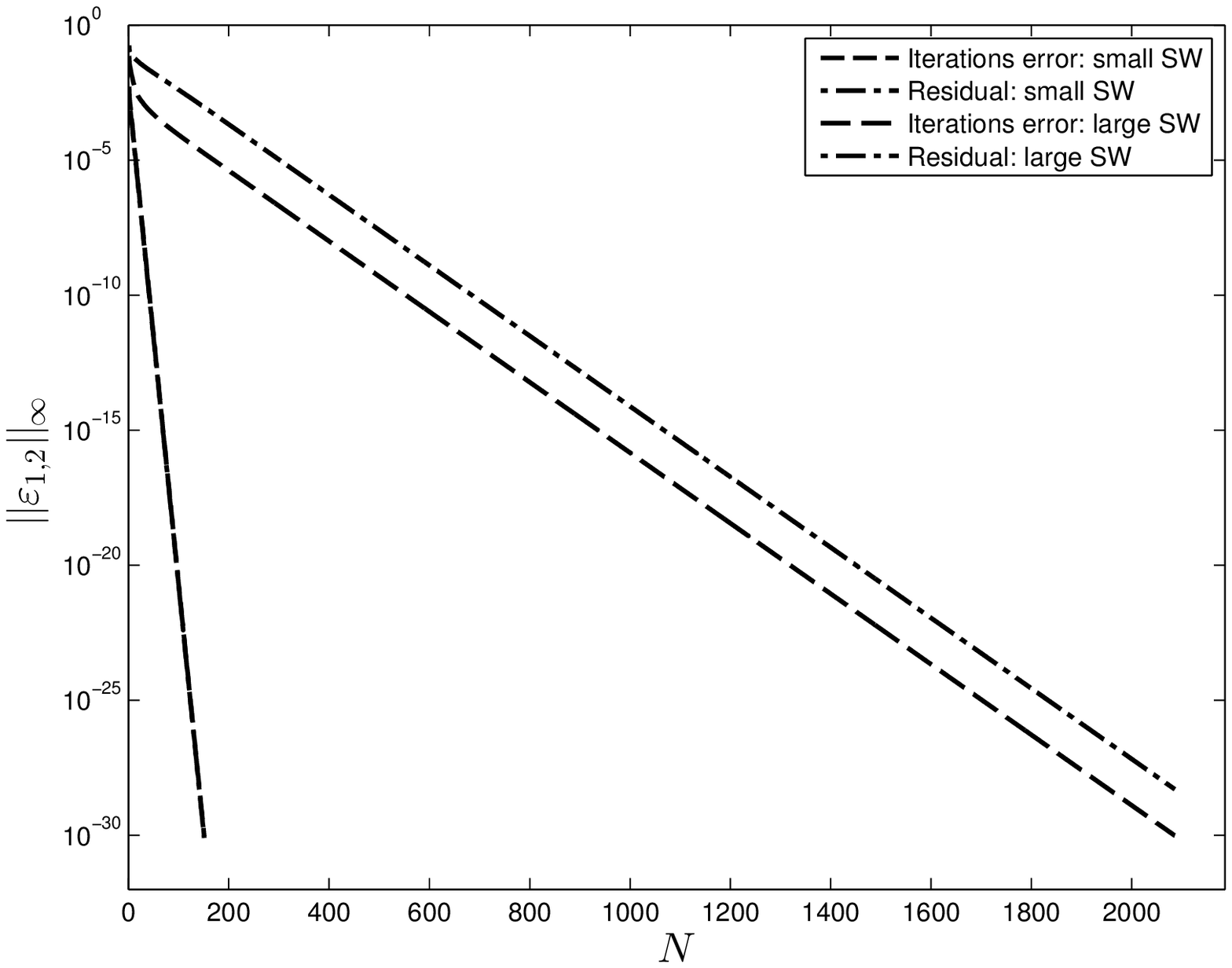}}
  \caption{\small\em Two solitary waves computed in multiprecision arithmetics along with the 
  convergence curves showing the decay of the error in $L_\infty$ norm.}
  \label{fig:mp}
\end{figure}

Now, we take this Froude number and use it as the solitary wave parameter in the Babenko equation \eqref{eq:babpet}, which is posed on the 1D periodic domain which is chosen adaptively so that, at the end points of this interval, the solitary wave tail drops below the machine accuracy. The interval is discretised using $N=16,384$ Fourier modes. The tolerance is chosen to be $10^{-15}$, i.e. close to machine accuracy. These parameters will be kept for all computations presented below. The iterative process is stopped when the $L_\infty$ norm of the difference between two successive iterations is less than the tolerance (the norm of the residual is checked a posteriori). Our algorithm required $70$ 
iterations to fully converge and the computations lasted slightly less than$0.01\, \textsf{s}$. The resulting solution is shown on Figure~\ref{fig:mp} (left, dashed line). This flagrant difference in CPU times can be explained by two main factors. On the one hand, the convergence rate of the Petviashvili scheme is slightly higher --- $70$ iterations compared to $173$ iterations for the Tanaka method. On the other hand, each iteration of our method has the super-linear complexity $\O(N\log N)$ while Tanaka's method, which relies on the computation of some singular integrals, has a quadratic complexity $\O(N^2)$. Now, let us compare the amplitudes of solitary waves:
\begin{align*}
a_{\mbox{\tiny\sc Tan}}\,/\,d\ =\ \textbf{0.1382189}0,\qquad
a_{\mbox{\tiny\sc Pet}}\,/\,d\ =\ \textbf{0.1382189}387245721,
\end{align*}
where $a_{\mbox{\tiny\sc Tan}}$ and $a_{\mbox{\tiny\sc Pet}}$ stand for the amplitudes predicted 
by Tanaka's and Petviashvili's methods, respectively. One can see that the first seven digits agree. We recall that the tolerance in Tanaka's algorithm was set to $10^{-10}$. So, the numerical results are somehow coherent with the prescribed numerical parameters. Solitary waves shapes are compared on Figure~\ref{fig:small}(a). However, the two solutions cannot be distinguished to the graphical resolution. That is why we plotted also on Figure~\ref{fig:small}(b) the difference between the curves. The discrepancies are of the order of $10^{-7}$ in agreement with the comparison of the amplitudes.

\begin{figure}
  \centering
  \subfigure[Velocity potential]{%
  \includegraphics[width=0.49\textwidth]{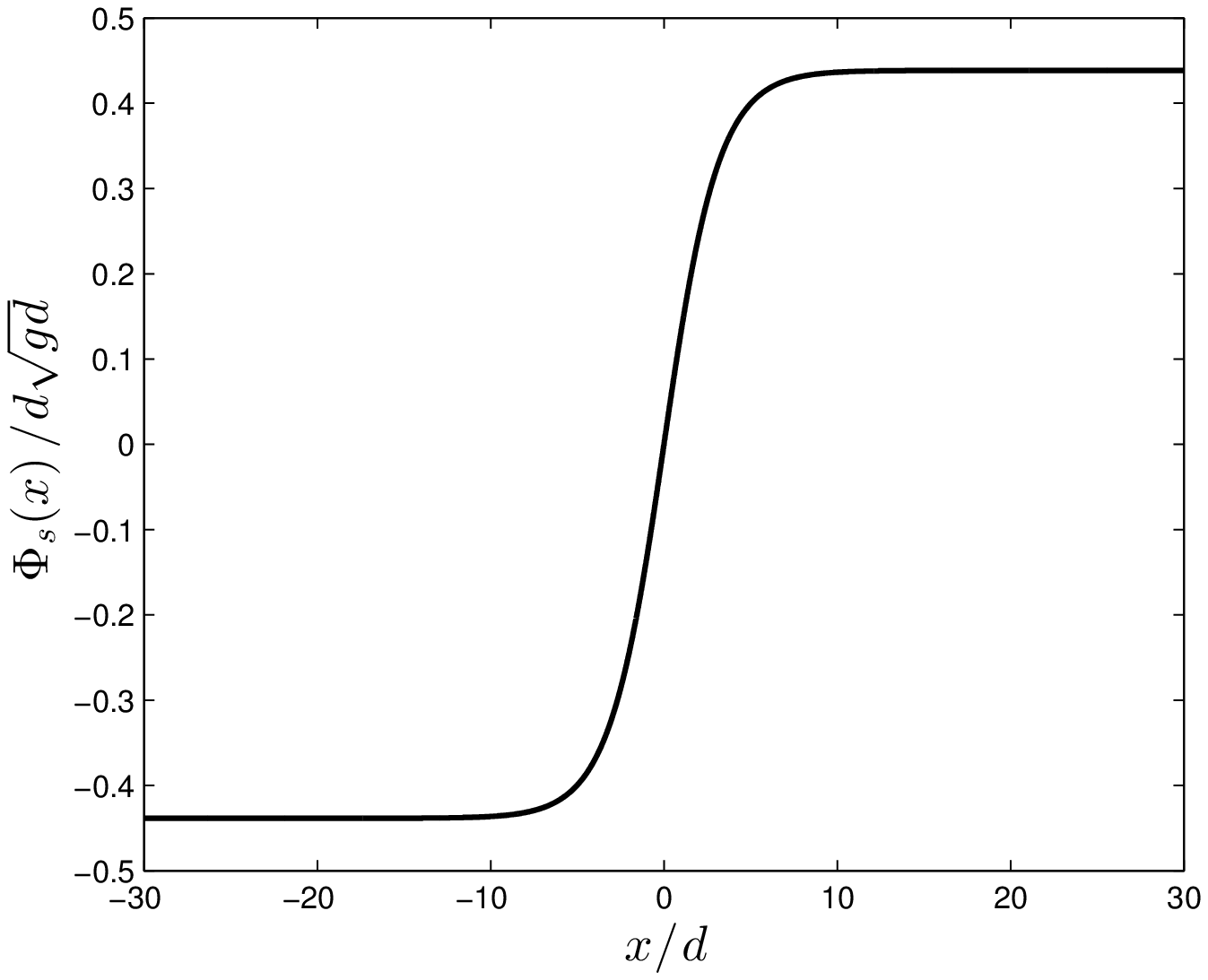}}
  \subfigure[Stream function]{%
  \includegraphics[width=0.49\textwidth]{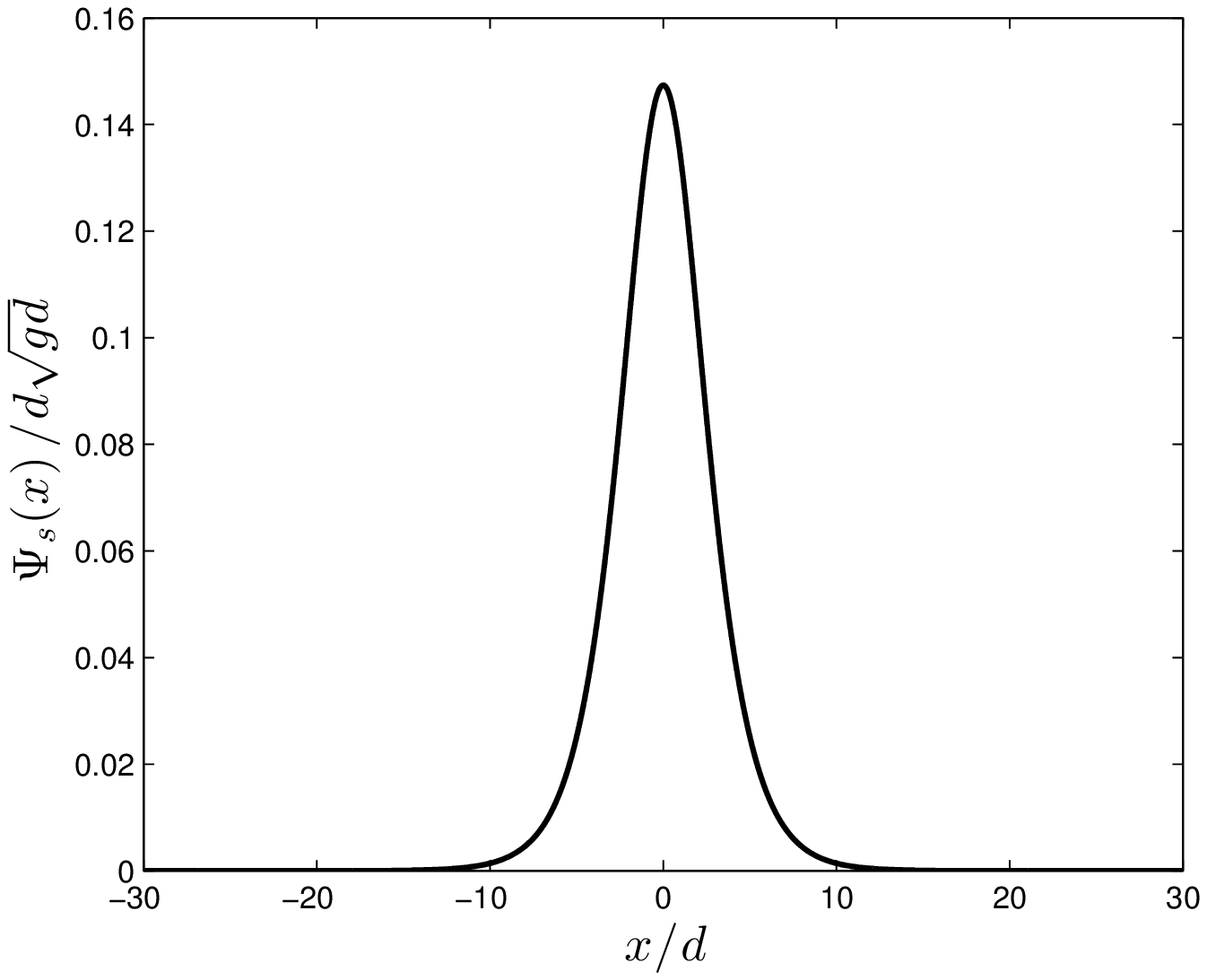}}
  \subfigure[Horizontal velocity]{%
  \includegraphics[width=0.49\textwidth]{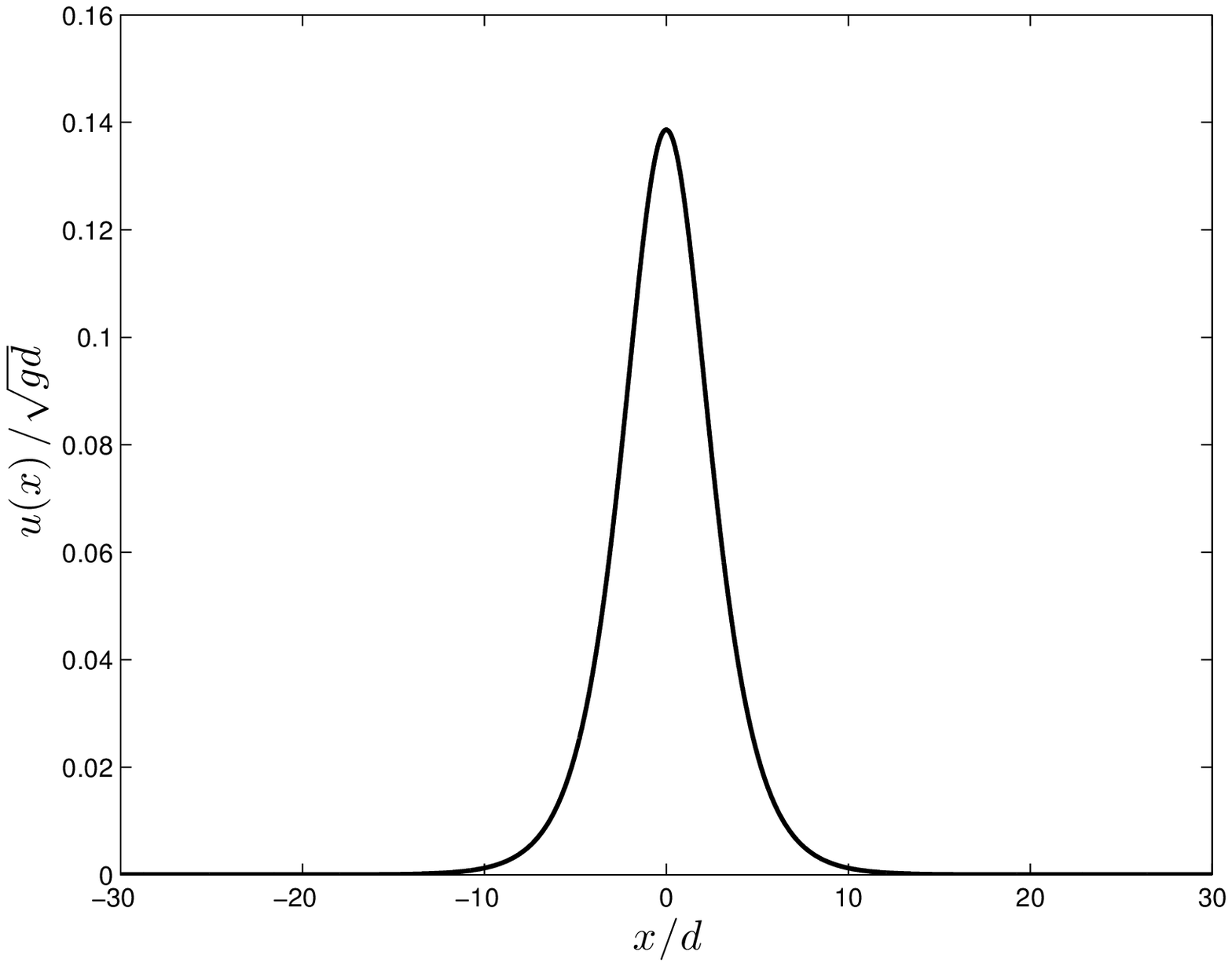}}
  \subfigure[Vertical velocity]{%
  \includegraphics[width=0.49\textwidth]{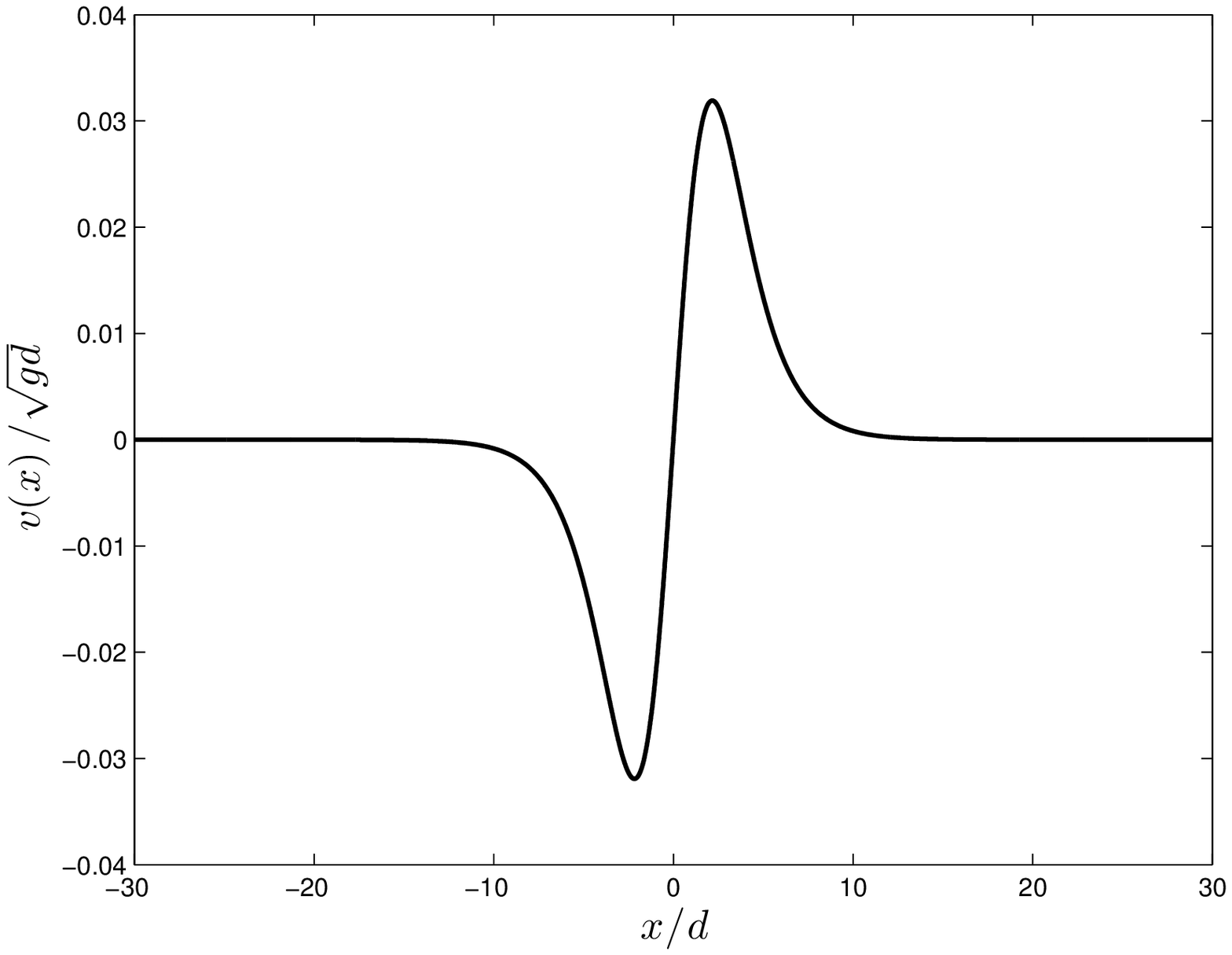}}
  \caption{\em\small The velocity potential, stream function, horizontal and vertical velocities evaluated at the free surface for the Froude number $F_{\mbox{\tiny\sc Tan}} = 1.066365888477383$. (Note the different scales between the plots.)}
  \label{fig:pot}
\end{figure}

Another feature of the conformal mapping consists in the easiness to obtain the velocity 
potential at the free surface along with the stream function needed, in particular, 
for transient computations \cite{Craig1993, Fructus2005}:
\begin{equation*}
\sur{\Phi}(x)\ \equiv\ \Re\/\{\/\Xi(x+\ui\eta)\/\}\ =\ c\,X(\alpha,0), \qquad
\sur{\Psi}(x)\ \equiv\ \Im\/\{\/\Xi(x+\ui\eta)\/\}\ =\ c\,Y(\alpha,0).
\end{equation*}
The graphs of the velocity potential and the stream function computed this way are plotted on Figure~\ref{fig:pot}(a,b). We underline that these quantities can be obtained just by applying a few simple post-processing operations to the converged Babenko equation solution. By using similar explicit representation one can also compute the horizontal and vertical velocities computed at the free surface. They are represented for illustrative purposes on Figure~\ref{fig:pot} correspondingly (the solitary wave is the same for all plots on Figure~\ref{fig:pot}).

Now, let us compare these numerical solutions with the available analytical results. Since we are dealing with a small amplitude solitary wave, we can apply the ninth-order Fenton asymptotic expansion \cite{Fenton1972} for the solitary wave speed $c^2$ in terms of its nonlinearity $\eps=a/d$. So, we take the amplitude $a_{\mbox{\tiny\sc Pet}}/d = 0.1382189387245723$ computed from the Babenko equation and, using Fenton's expansion, we estimate the solitary wave speed for this amplitude:
\begin{align*}
c_{\mbox{\tiny\sc Fen}}\,/\,\sqrt{gd}\ =\ \textbf{1.066365888}5, \qquad
c_{\mbox{\tiny\sc Pet}}\,/\,\sqrt{gd}\ =\ \textbf{1.066365888}477383,
\end{align*}
where $c_{\mbox{\tiny\sc Pet}}$ is the solitary wave speed parameter used in the Petviashvili scheme. One can see that the differences between the predicted (Fenton) and prescribed (Petviashvili) values start to appear after the ninth digit, which is in perfect agreement with the order of the asymptotic expansion. However, strictly speaking, the Fenton solution is valid only in the limit $a/d\to 0$. On the other hand, the Babenko equation can be used to compute much more nonlinear solitary wave to the arbitrary precision. To illustrate the last statement we implemented the Petviashvili scheme using the 
Multiprecision Computing (MC) Toolbox for \textsc{Matlab} \cite{MATLAB2012}. The transformations of the code needed to implement the arbitrary accuracy are really minimalistic. This constitutes one of the major advantages of the MC Toolbox.

In these higher-precision experiments, we take a periodic $\phi$-interval adapted for the increased accuracy ($30$ digits). Since the interval becomes longer, we have to increase also the number of Fourier modes up to $N = 32\,768$. All floating point operations are done with $30$ significant digits (plus $3$ control digits). The tolerance parameter is set to $10^{-30}$. First, we compute the small solitary wave from previous examples using the extended arithmetics. We can compare the amplitudes with the standard double accuracy
\begin{align*}
  a_{\mbox{\tiny\sc dp}}\,/\,d\ =&\ \textbf{0.138218938724572}1, \\
  a_{\mbox{\tiny\sc hp}}\,/\,d\ =&\ \textbf{0.138218938724572}45734239592556073752.
\end{align*}
One can see that the first 15 digits computed with the standard arithmetics are correct, which validates one more time the algorithm (we recall that the tolerance was set to $10^{-15}$ in the double precision computation).

\begin{figure}
  \subfigure[]%
  {\includegraphics[width=0.49\textwidth]{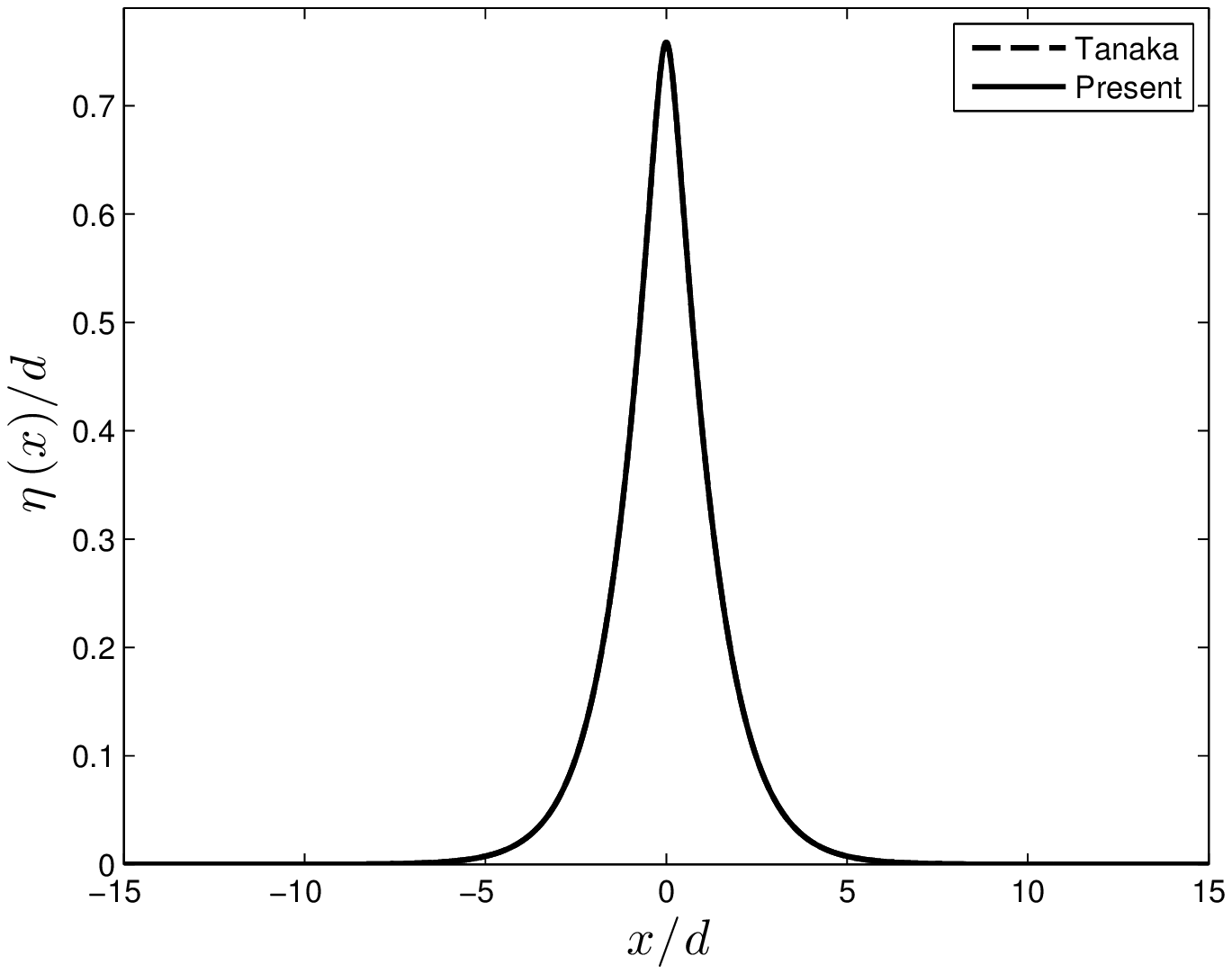}}
  \subfigure[]%
  {\includegraphics[width=0.49\textwidth]{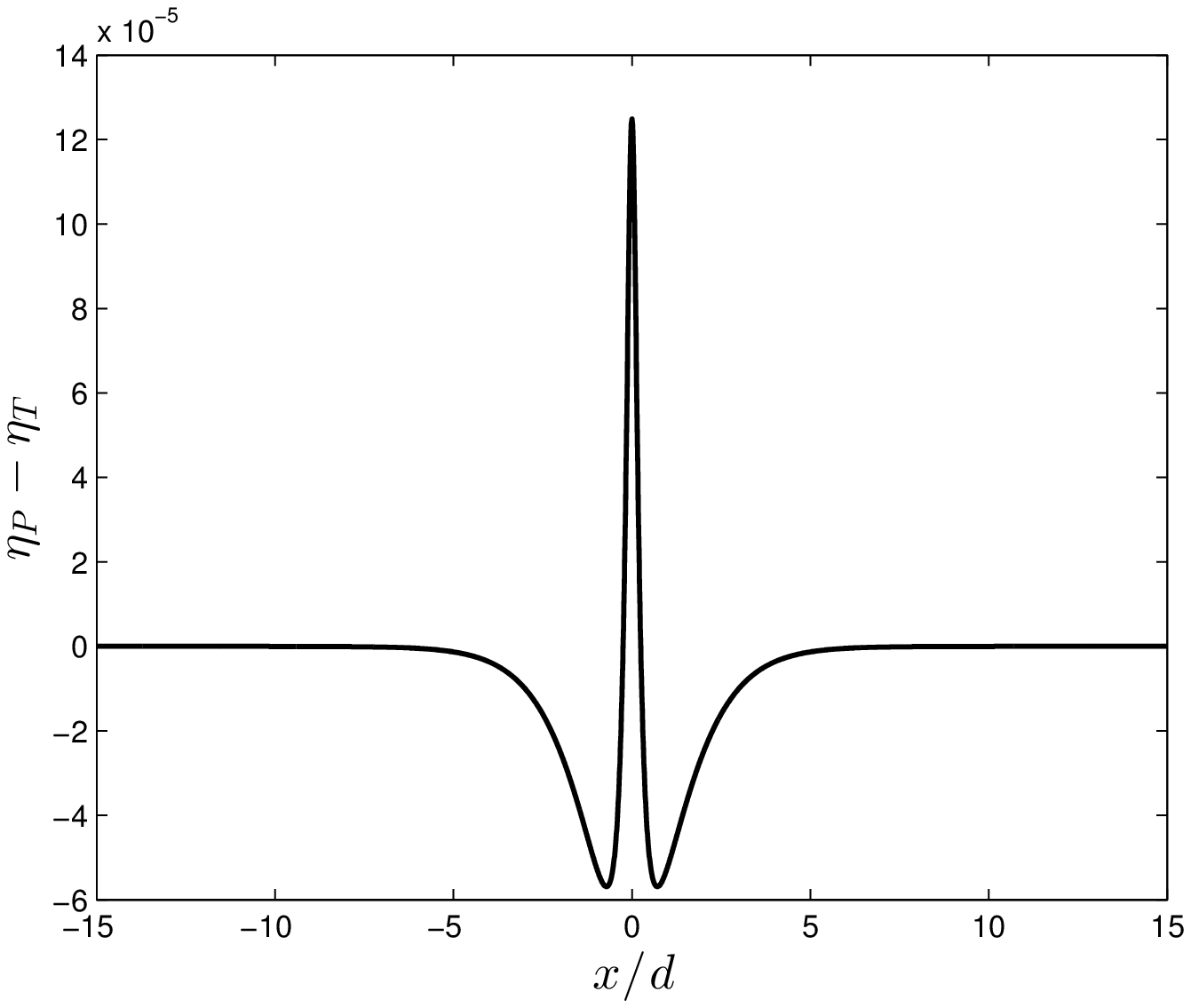}}
  \caption{\em\small Comparison between Tanaka's and the present solutions for the same Froude number $F_T = 1.290941713543984$. Left: free surface; Right: difference (the vertical scale is $10^{-5}$).}\label{fig:hi}
\end{figure}

However, it is much more challenging to compute high amplitude waves. The next example is inspired again by the Tanaka solution corresponding to the parameter $q_c = 0.3$. The dimensionless speed of propagation of this soliton is approximatively equal to
\begin{equation*}
  c\,/\,\sqrt{gd}\ =\ 1.290941713543984.
\end{equation*}
By fixing this parameter in the Babenko equation, we start the Petviashvili iterations. The result is depicted on Figure~\ref{fig:mp}(a) (solid line). The amplitude of this large solitary wave can be easily computed as well
\begin{equation*}
  a\,/\,d\ \approx\ 0.7583938551160400485984861886035\cdots.
\end{equation*}
However, this number is more difficult to validate, since in most studies the authors considered only solitary waves not higher than $a/d = 0.7$ (see, for example, \cite{Li2004}). Nevertheless, we performed a comparison with the Tanaka solution. The results are shown on Figure~\ref{fig:hi}. The maximal difference is of the order of $10^{-4}$ which is not completely satisfactory.

The computation of this large amplitude solitary wave took $2\,086$ iterations to converge to the tolerance $10^{-30}$ ($931$ iterations are needed to cross the standard $10^{-15}$ accuracy). The convergence rate of the iterative process is lower for high amplitude solitary waves. The $L_\infty$-norm of the difference between two successive iterations, along with the residual error are plotted on Figure~\ref{fig:mp}(b). One can notice that the convergence to the small wave is much faster, even if the process is still converging. This point could be probably improved in future studies applying, for example, the generalized Petviashvili iteration \cite{Lakoba2007}. Nevertheless, we have still a very good performance. For example, the computation of that large solitary wave to the standard $10^{-15}$ accuracy requires only $1.18\,${\s} (in terms of the CPU time) under \textsc{Matlab}. This timing is to be compared with Tanaka's results.

For the sake of efficiency, we shall execute hereinafter the computations only within the standard double precision floating point arithmetics \cite{Kahan1979}, unless the higher accuracy is explicitly required for some application.

\subsubsection{Conserved integral quantities}

Once the algorithm is validated, we can use it to produce some physically sound numerical results. The question we are addressing in this Section is the dependence of several important integral invariants (presented above in Section~\ref{sec:intq}) on the wave amplitude. This question has already been addressed in previous studies essentially using asymptotic methods \cite{Longuet-Higgins1974}, which are formally valid only for the small amplitude waves. Our approach does not have such limitations and we are going to apply it to explore the whole range of wave heights.

\begin{figure}
  \centering
  \subfigure[]{\includegraphics[width=0.49\textwidth]{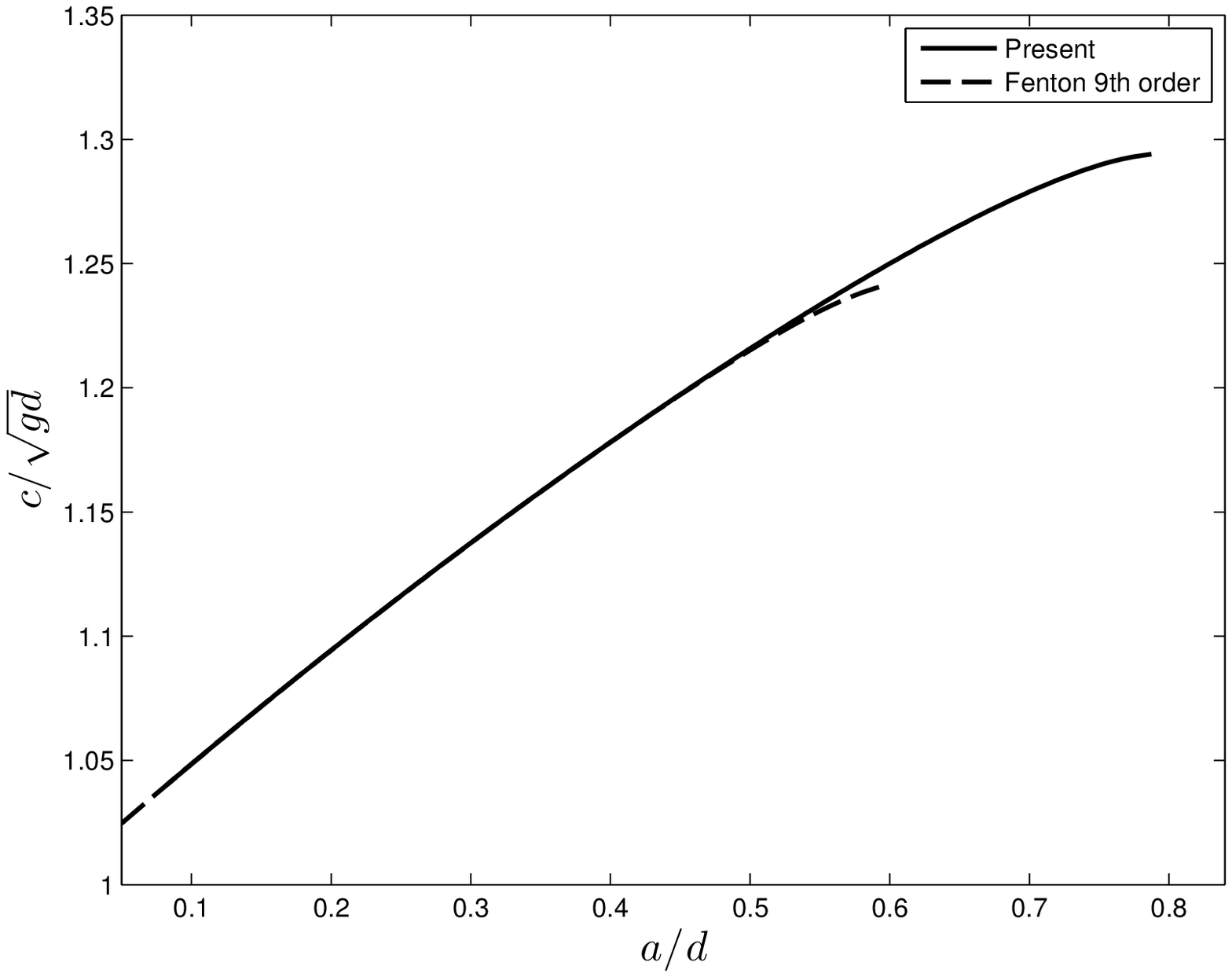}}
  \subfigure[]{\includegraphics[width=0.49\textwidth]{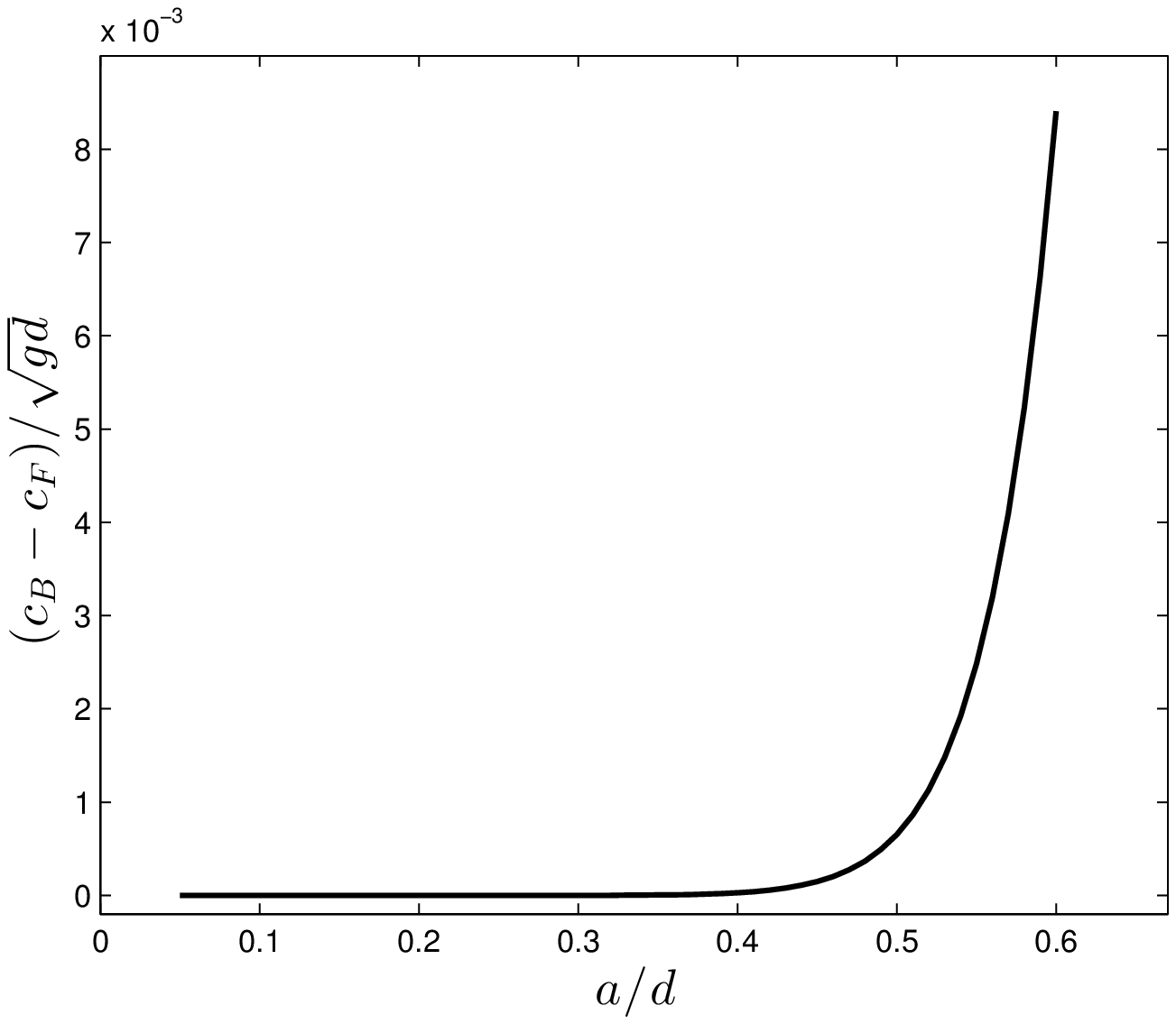}}
  \caption{\em (a) Speed-amplitude relation for solitary gravity waves to the full Euler equations. Comparison with the Fenton ninth-order expansion. (b) Difference between Babenko and Fenton predictions for the speed-amplitude relations (the vertical scale on the right image (b) is $10^{-3}$).}
  \label{fig:csa}
\end{figure}

\begin{figure}
  \centering
  \subfigure[Mass $\mathcal{M}$]{%
  \includegraphics[width=0.49\textwidth]{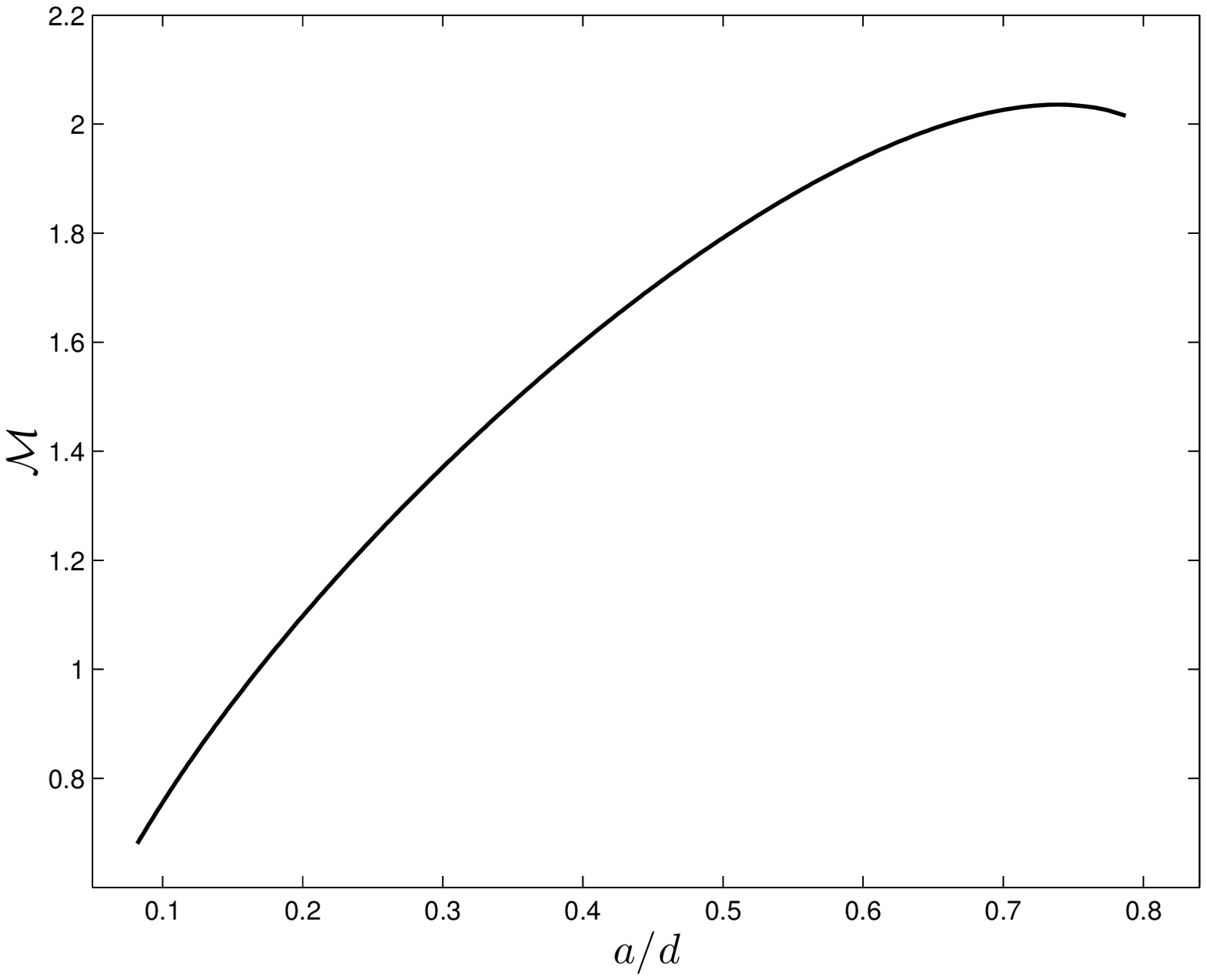}}
  \subfigure[Circulation $\mathcal{C}$]{%
  \includegraphics[width=0.49\textwidth]{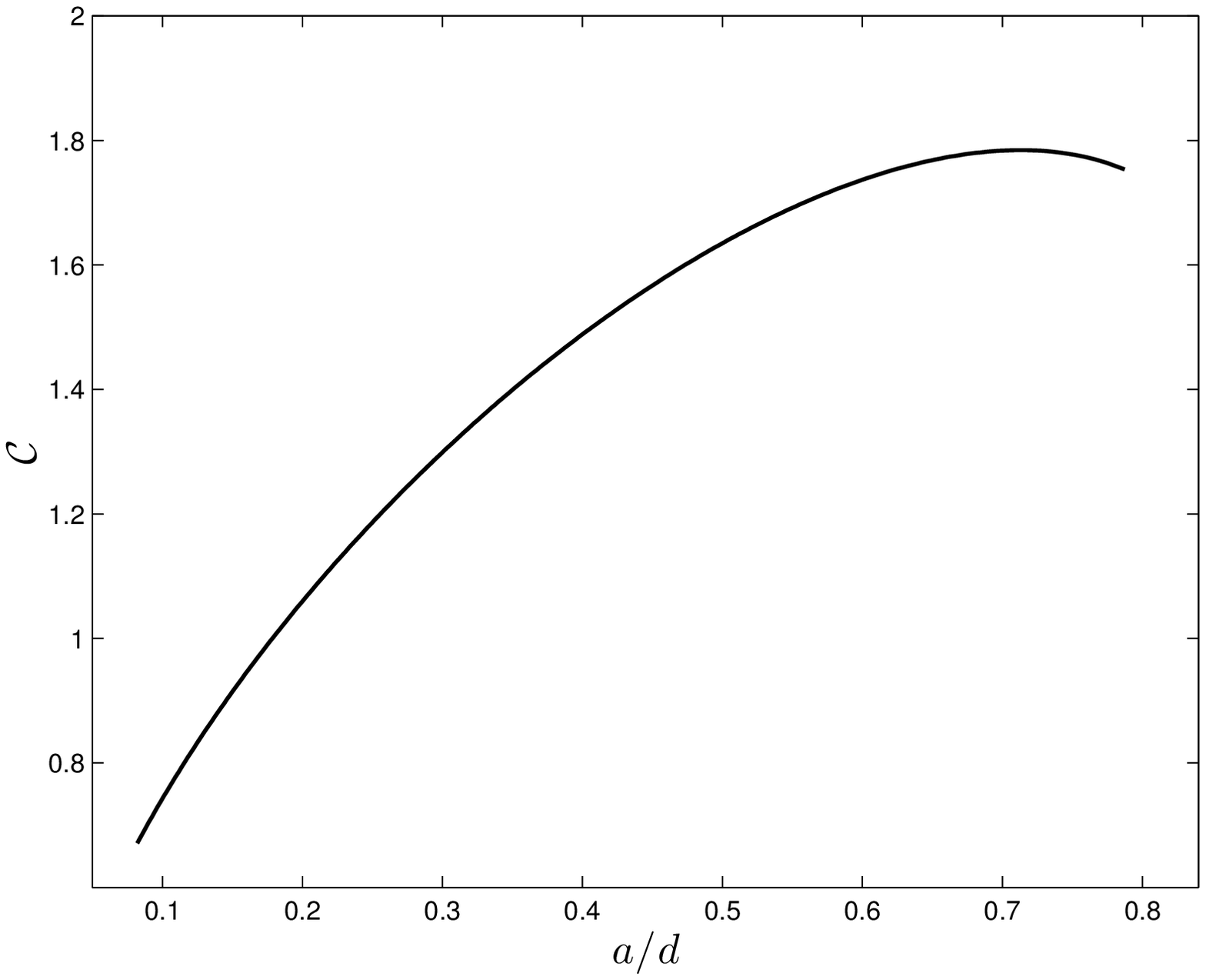}}
  
  \subfigure[Kinetic energy $\mathcal{K}$]{%
  \includegraphics[width=0.49\textwidth]{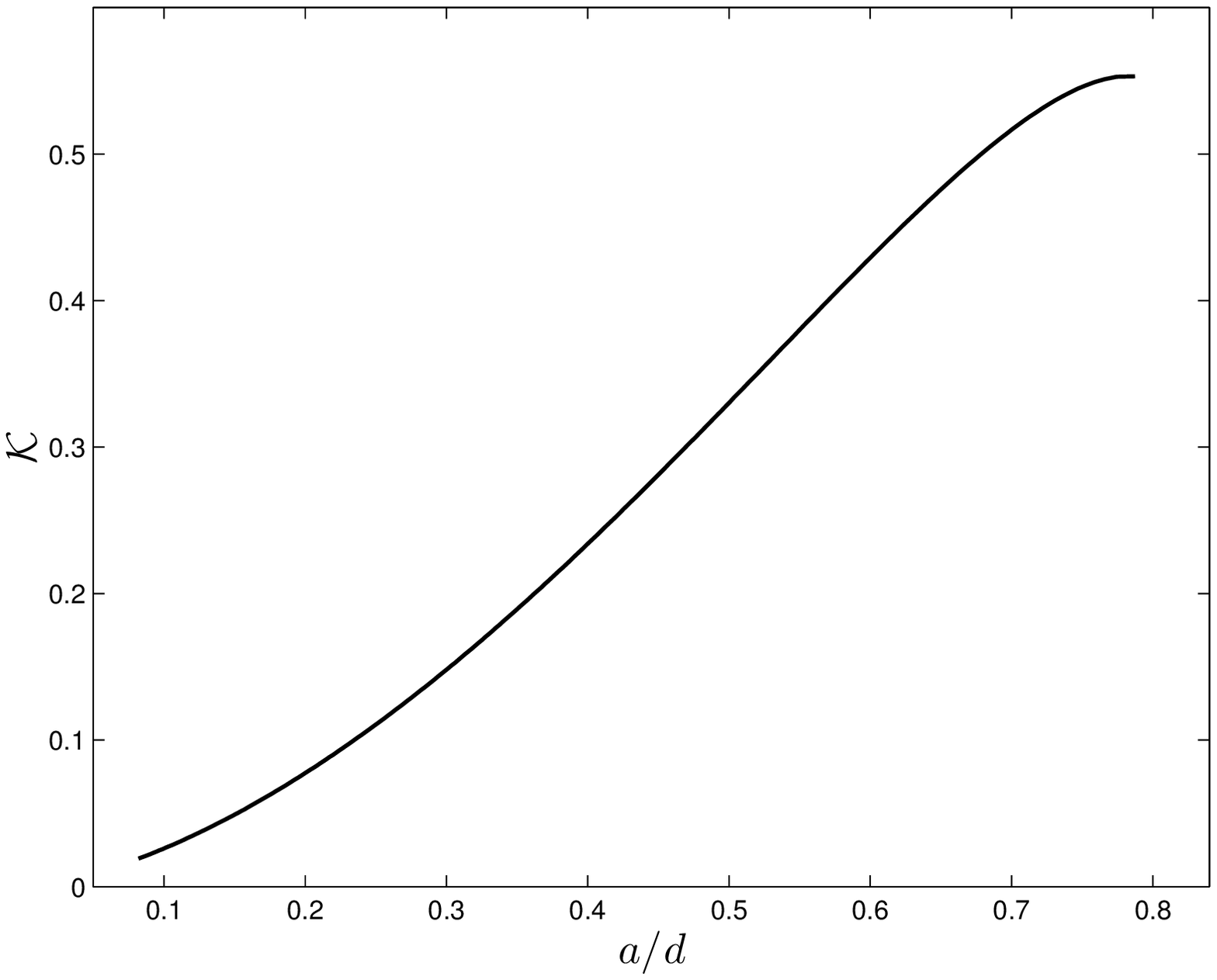}}
  \subfigure[Potential energy $\mathcal{V}_g$]{%
  \includegraphics[width=0.49\textwidth]{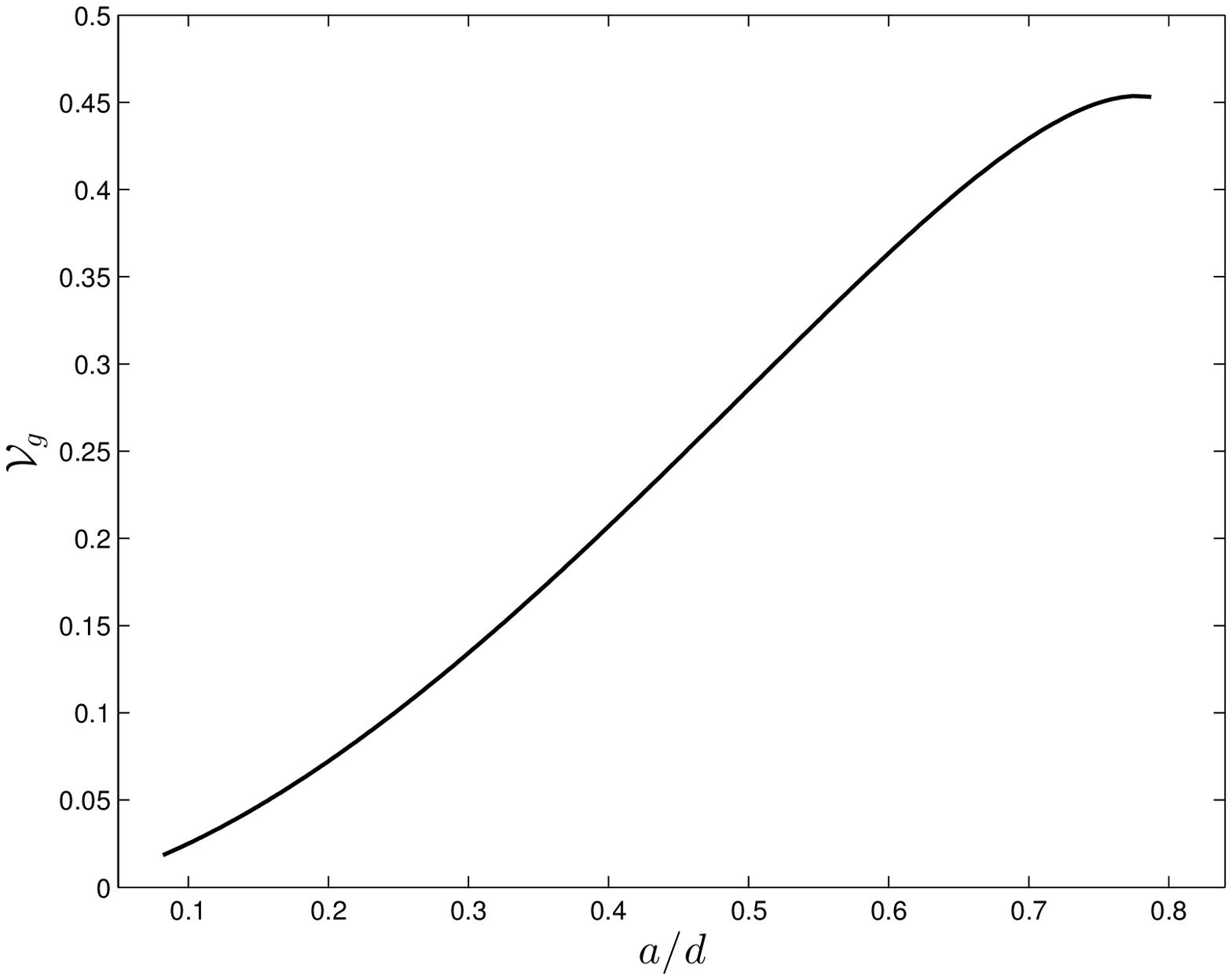}}
  
  \subfigure[Impulse $\mathcal{I}$]{%
  \includegraphics[width=0.49\textwidth]{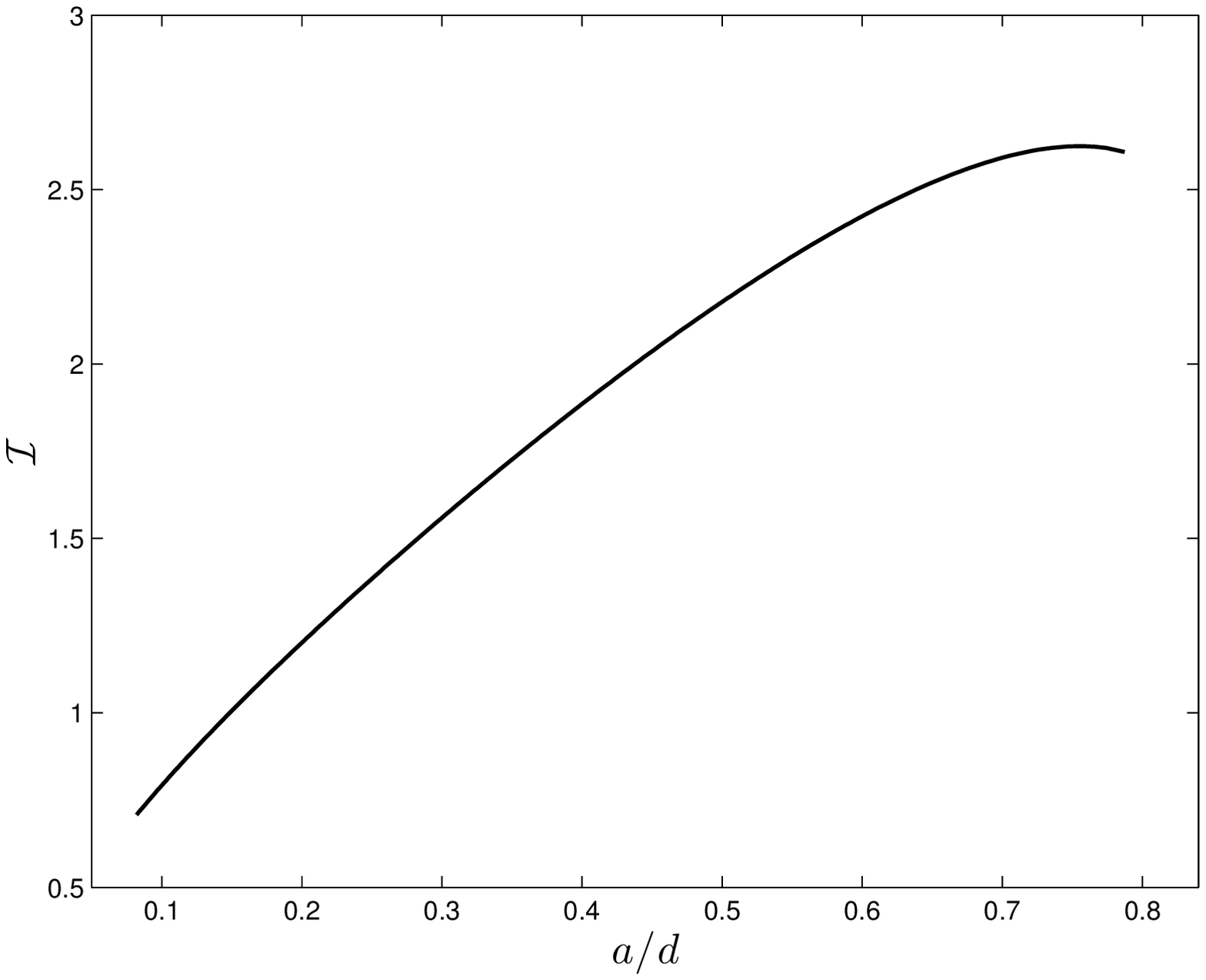}}
  \subfigure[Total energy $\mathcal{E}$]{%
  \includegraphics[width=0.49\textwidth]{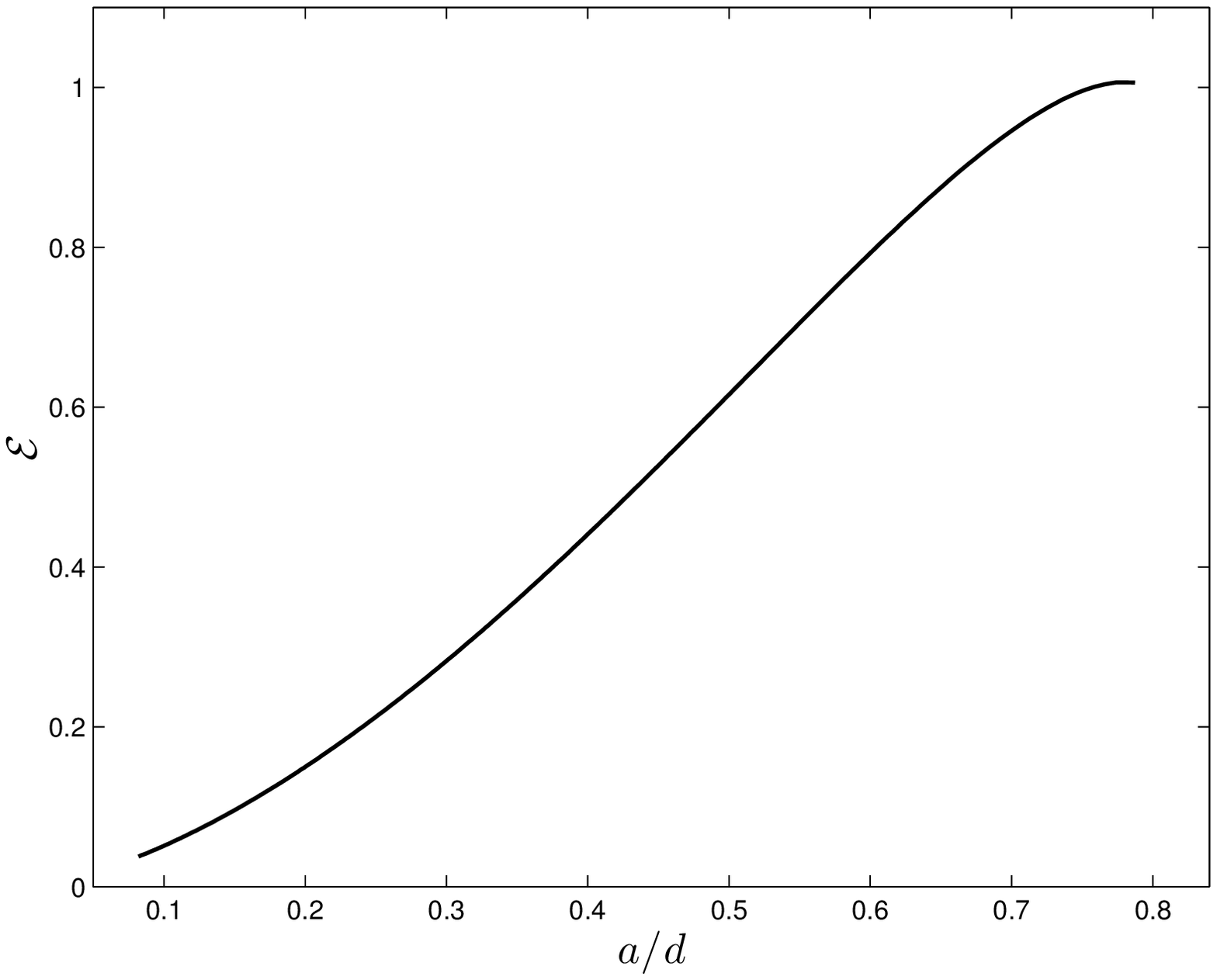}}
  
  \caption{\small\em The dependence of several integral quantities for solitary waves on the wave amplitude.}
  \label{fig:invs}
\end{figure}

First of all, on Figure~\ref{fig:csa} we present the so-called speed-amplitude relation --- the abscissa represents the dimensionless amplitude of the wave, while the vertical axis shows the corresponding propagation speed. Our numerical result is compared to the classical Fenton ninth-order solution \cite{Fenton1972}. Again, from small to moderate solitary waves we obtain a very good agreement. For large amplitudes the advantage of our method becomes more explicit. Finally, on Figure~\ref{fig:invs}(a)--(f) we show the dependence of the wave mass $\mathcal{M}$, circulation $\mathcal{C}$, kinetic energy $\mathcal{K}$, potential energy $\mathcal{V}_g$, impulse $\mathcal{I}$ and the total energy $\mathcal{E}$ correspondingly on the wave amplitude $a/d$. For instance, one can 
see that these quantities do not necessarily have the monotonic behaviour when the amplitude grows.

\begin{figure}
  \centering
  \subfigure[Potential]%
  {\includegraphics[width=0.49\textwidth]{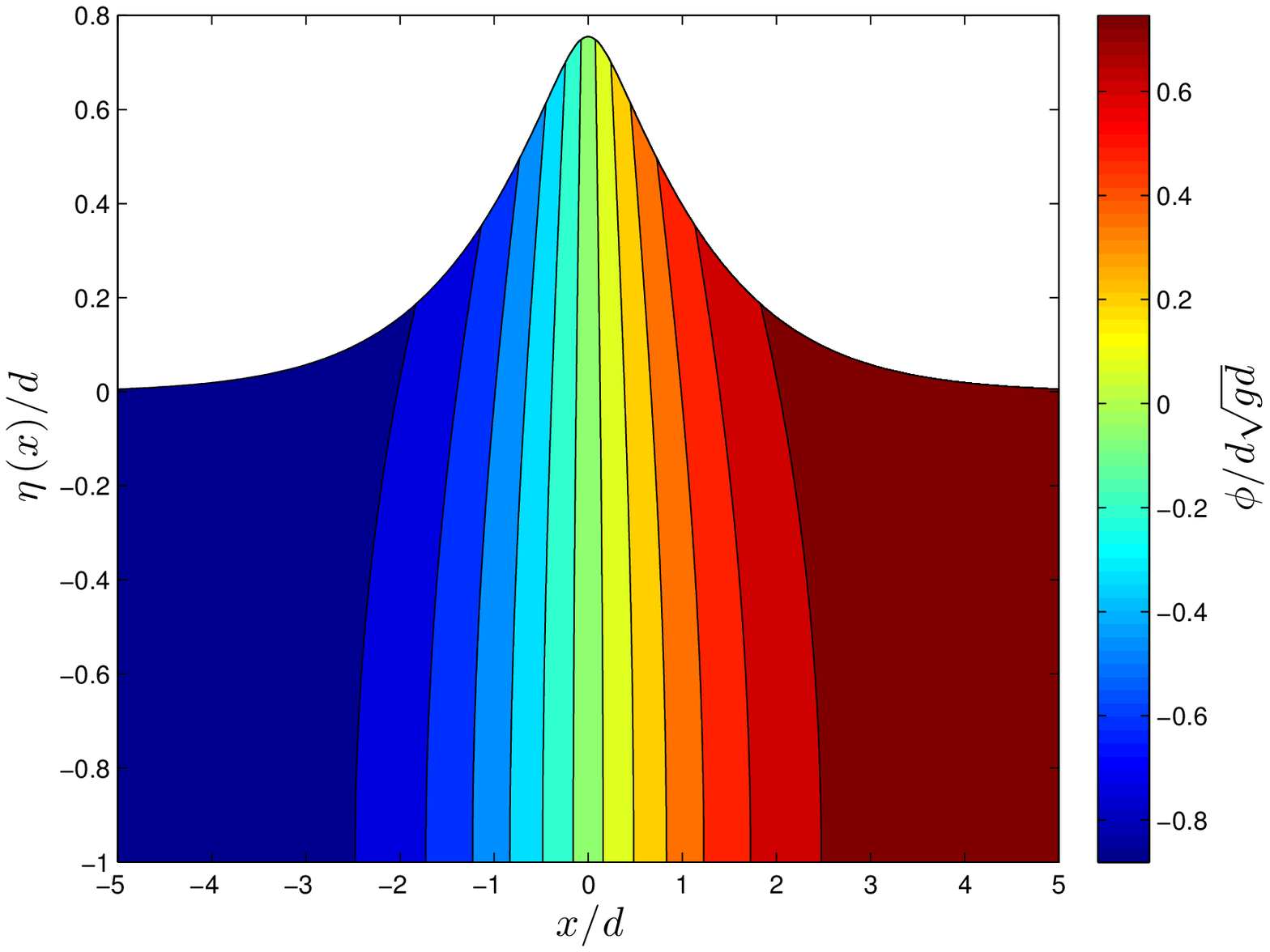}}
  \subfigure[Stream function]%
  {\includegraphics[width=0.49\textwidth]{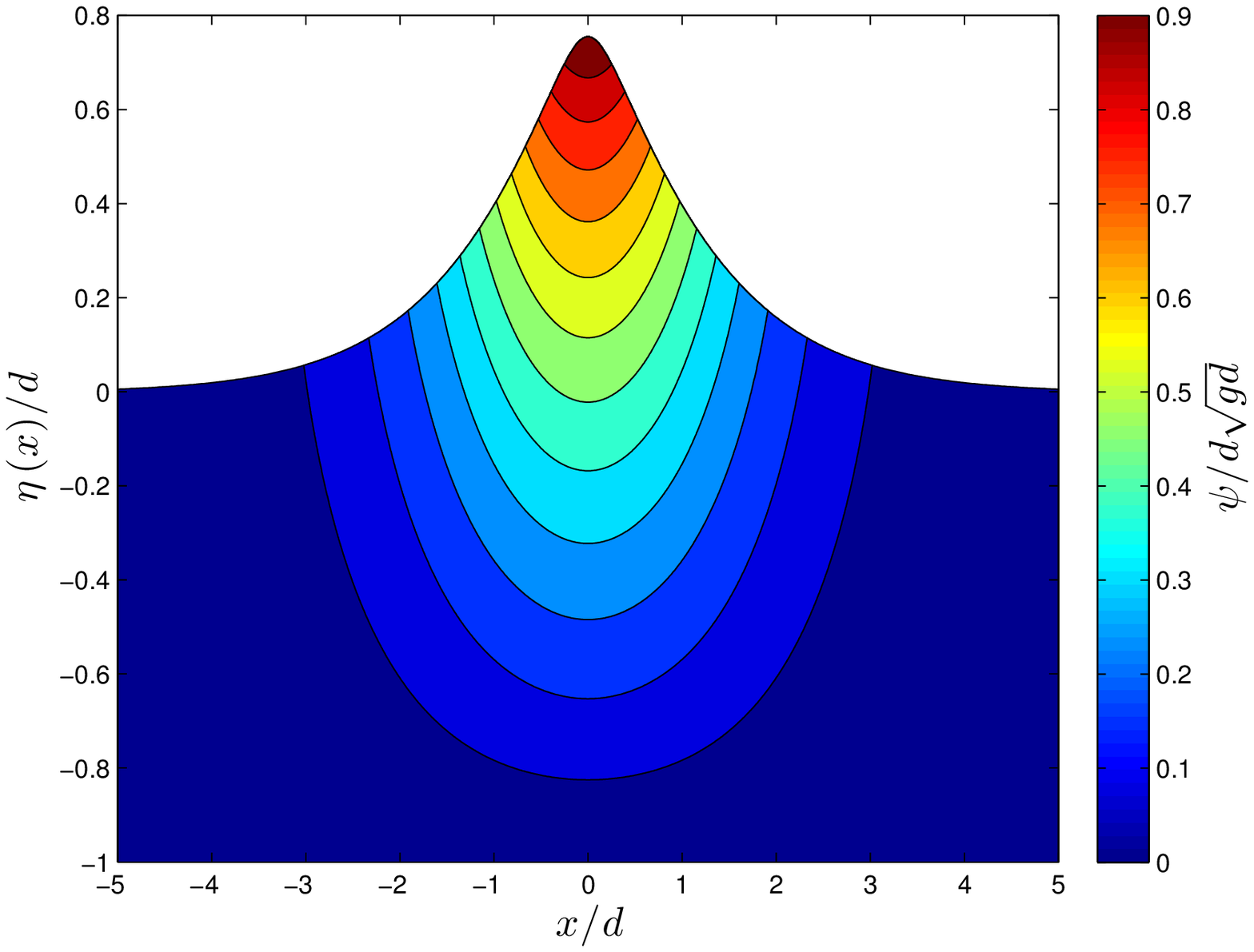}}
  \caption{{\small {\em Equi-potentials (left) and iso-stream-function (right) under a large wave. Lines correspond to the iso-values computed in the `fixed' Frame of reference where the the fluid is at rest in the far field $x\to\pm\infty$.}}}
  \label{fig:potstream}
\end{figure}

\begin{figure}
  \centering
  \subfigure[Total pressure]%
  {\includegraphics[width=0.49\textwidth]{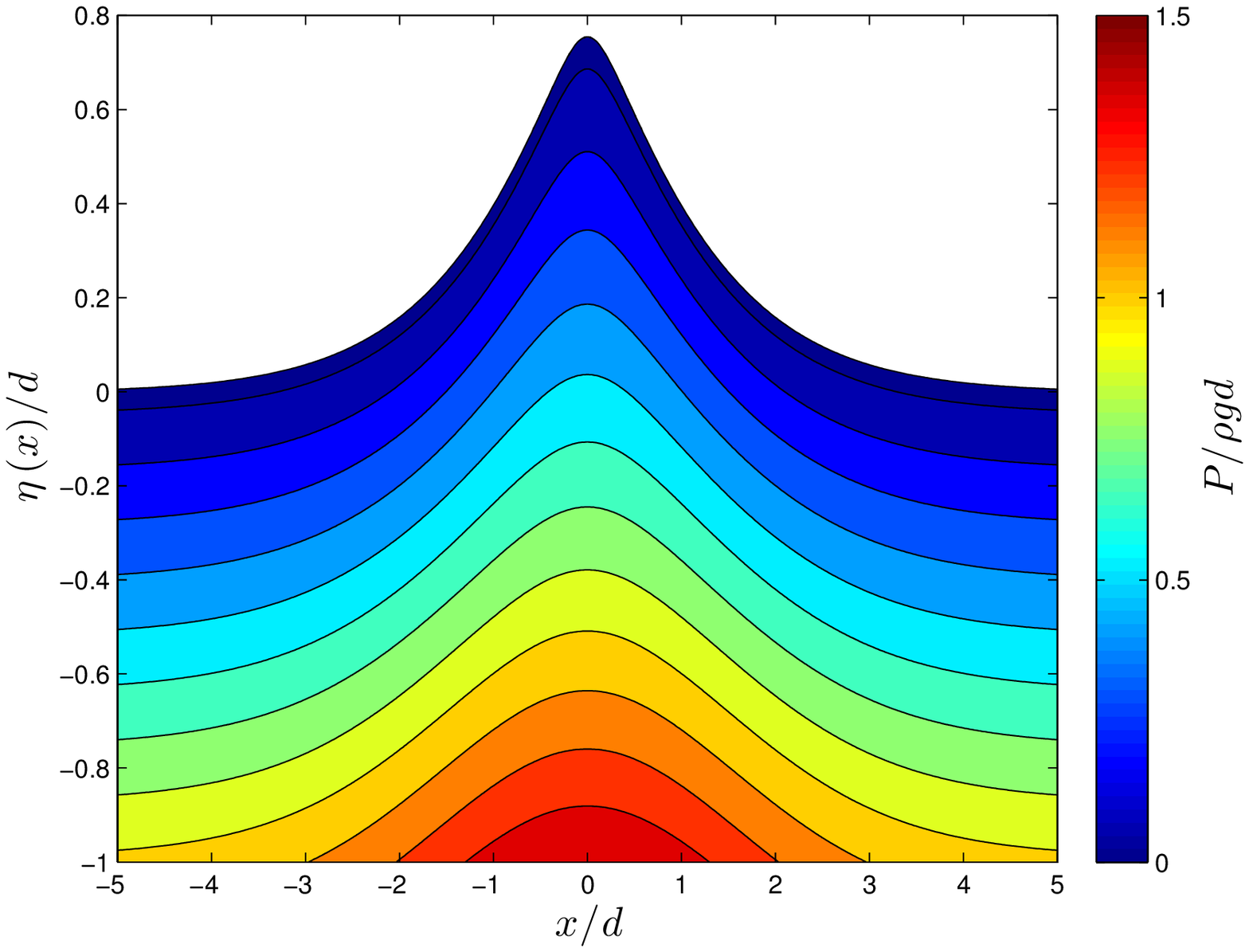}}
  \subfigure[Dynamic pressure]%
  {\includegraphics[width=0.49\textwidth]{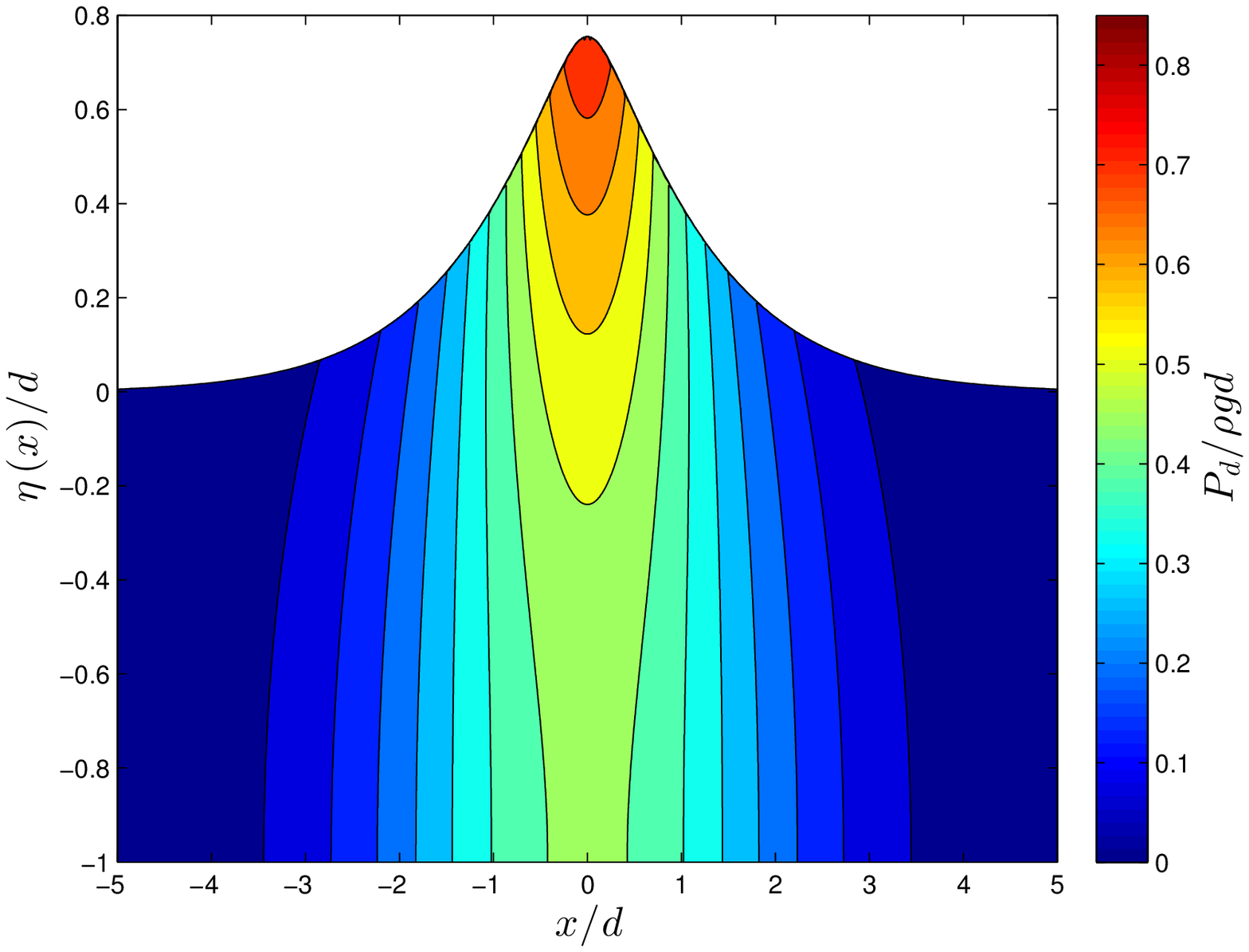}}
  \caption{\small\em Isobars (left) and iso-dynamic-pressures $p + gy$ (right) under a large wave.}
  \label{fig:press}
\end{figure}

\begin{figure}
  \centering
  \subfigure[Horizontal velocity]%
  {\includegraphics[width=0.49\textwidth]{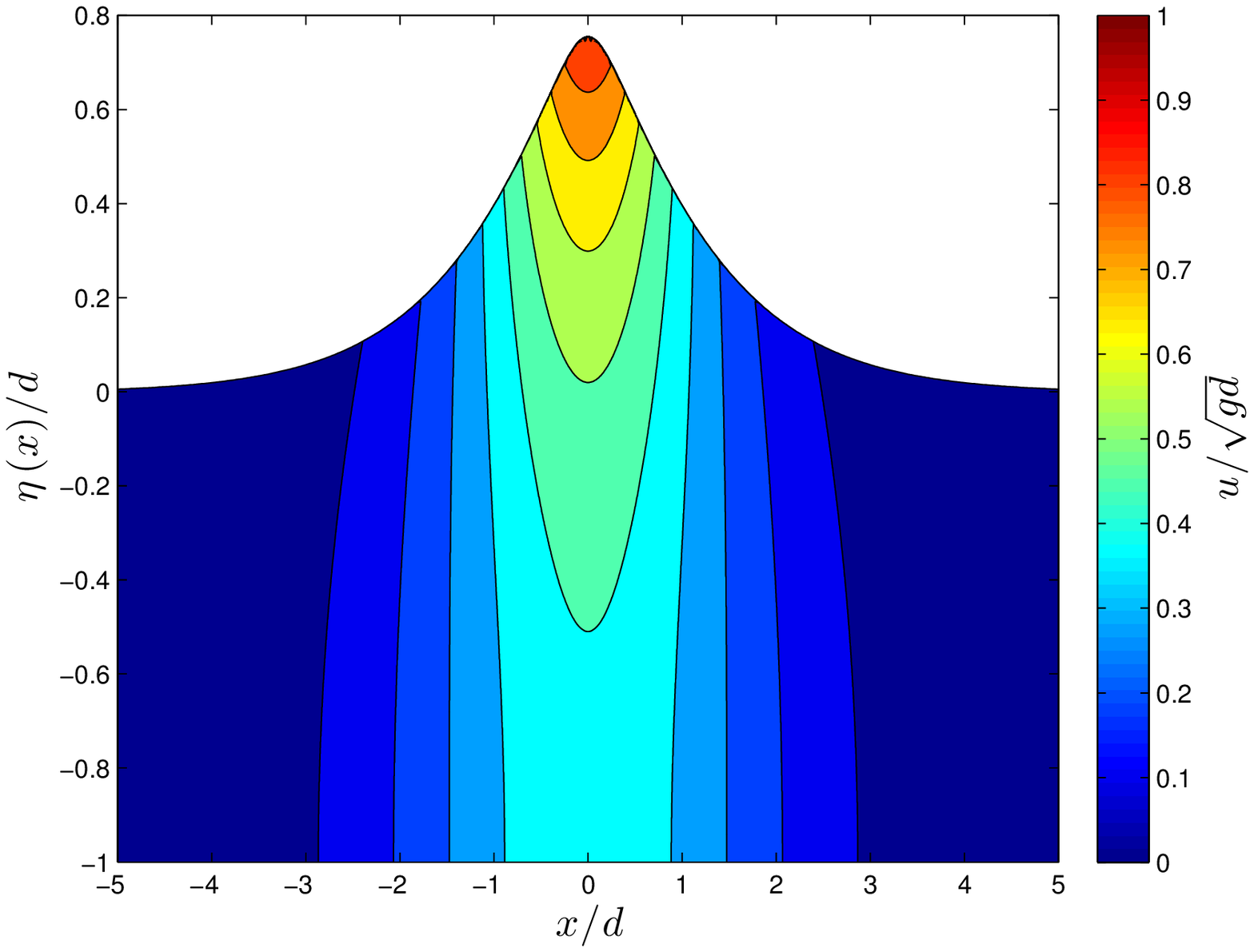}}
  \subfigure[Vertical velocity]%
  {\includegraphics[width=0.49\textwidth]{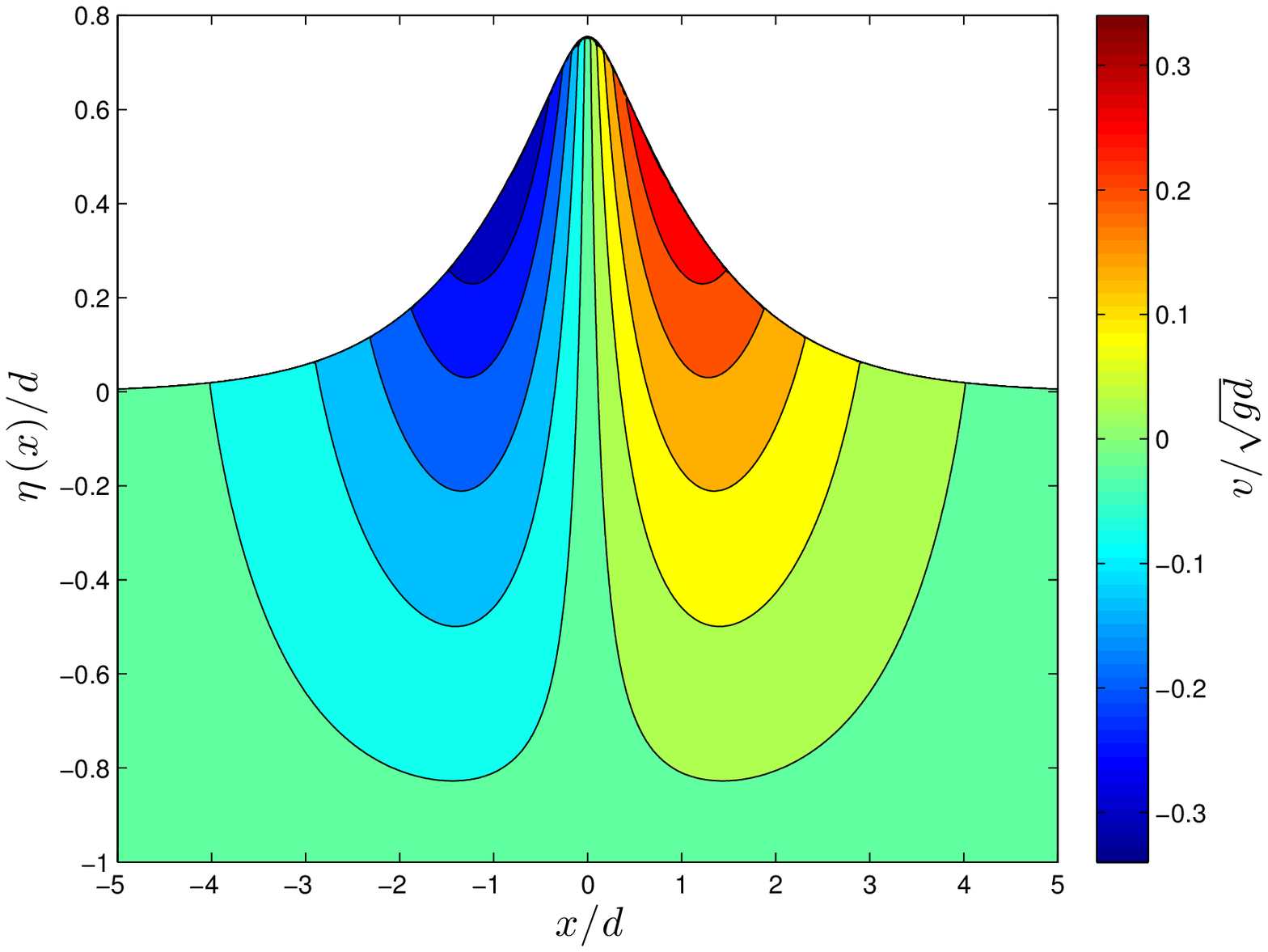}}
  \caption{\small\em Iso-horizontal (left) and iso-vertical (right) velocities under a large wave. Lines correspond to the iso-values computed in the `fixed' Frame of reference where the the fluid is at rest in the far field $x\to\pm\infty$.}
  \label{fig:speeds}
\end{figure}

\begin{figure}
  \centering
  \subfigure[Horizontal acceleration]%
  {\includegraphics[width=0.49\textwidth]{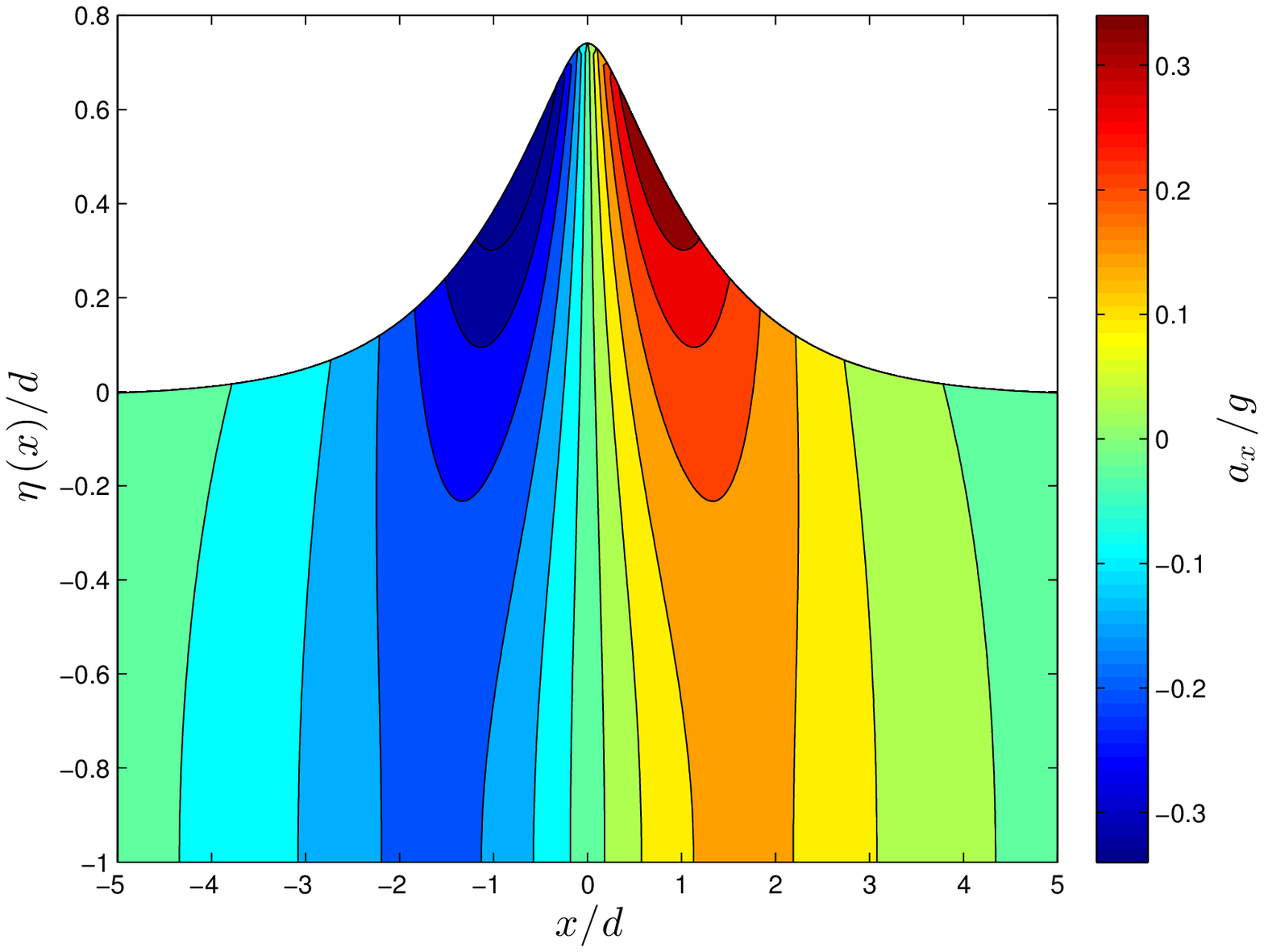}}
  \subfigure[Vertical acceleration]%
  {\includegraphics[width=0.49\textwidth]{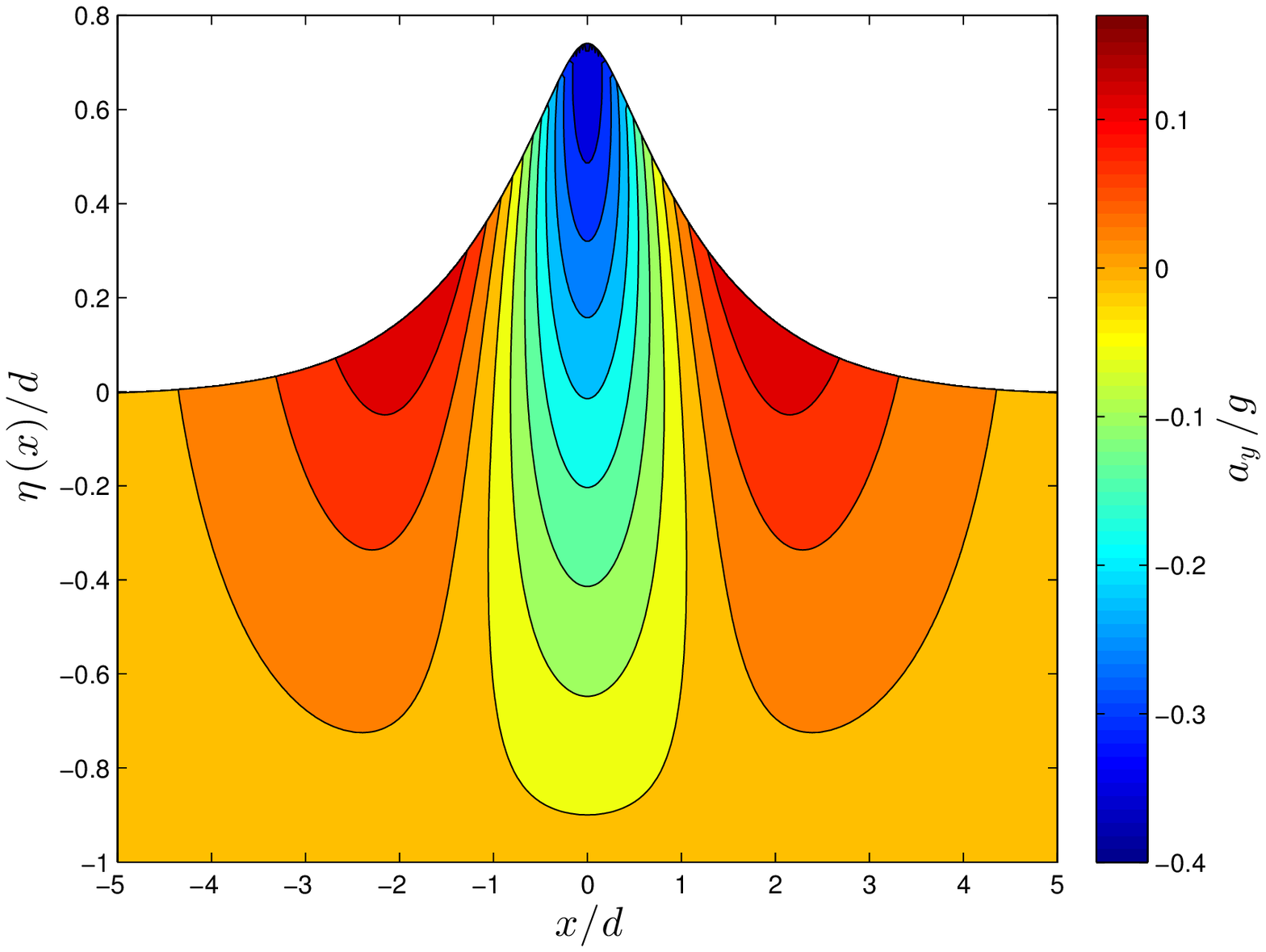}}
  \caption{\small\em Iso-horizontal (left) and iso-vertical (right) accelerations under a large wave.}
  \label{fig:accel}
\end{figure}

\begin{figure}
  \centering
  \subfigure[Kinetic energy density]%
  {\includegraphics[width=0.49\textwidth]{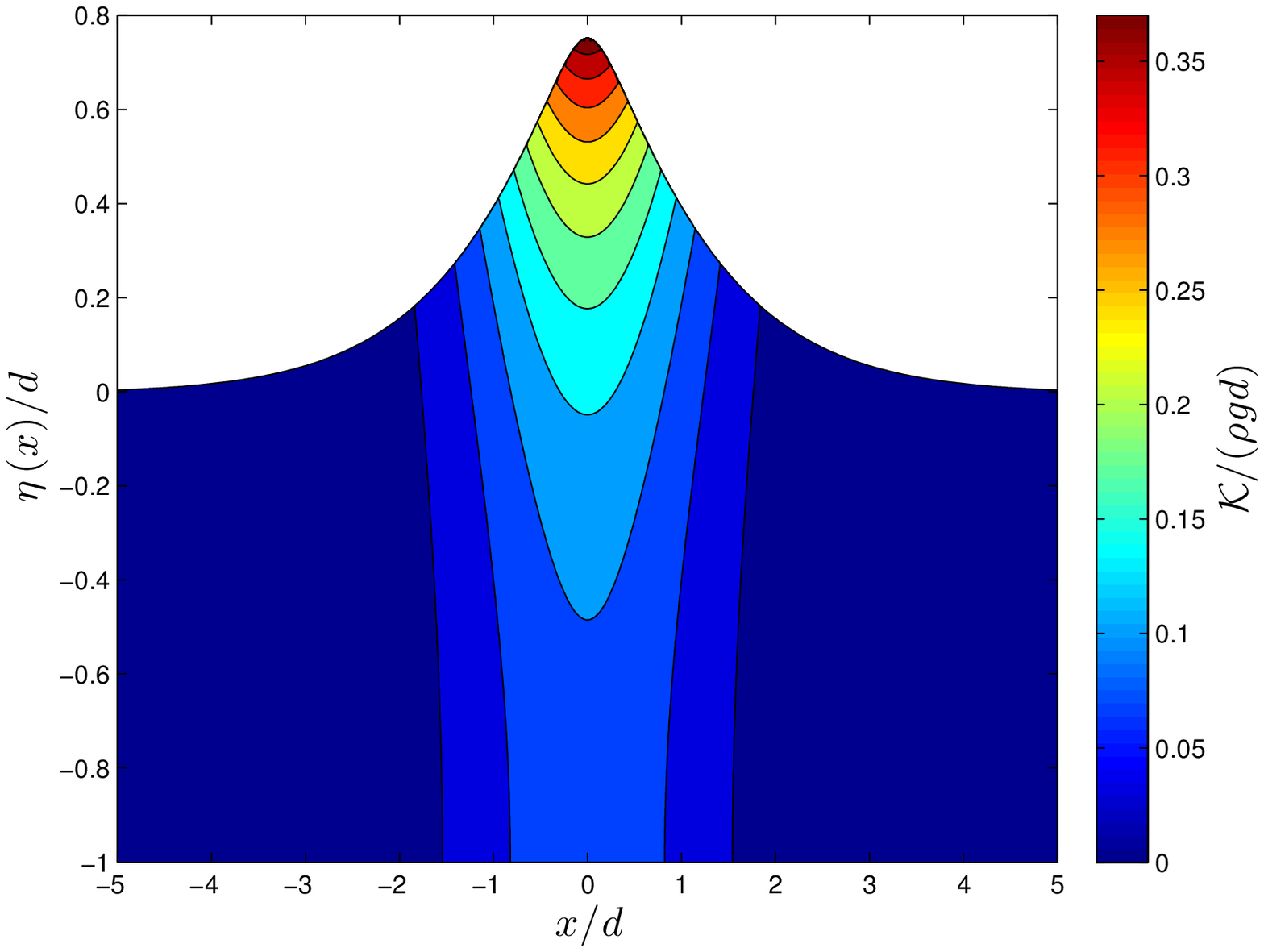}}
  \subfigure[Total energy flux]%
  {\includegraphics[width=0.49\textwidth]{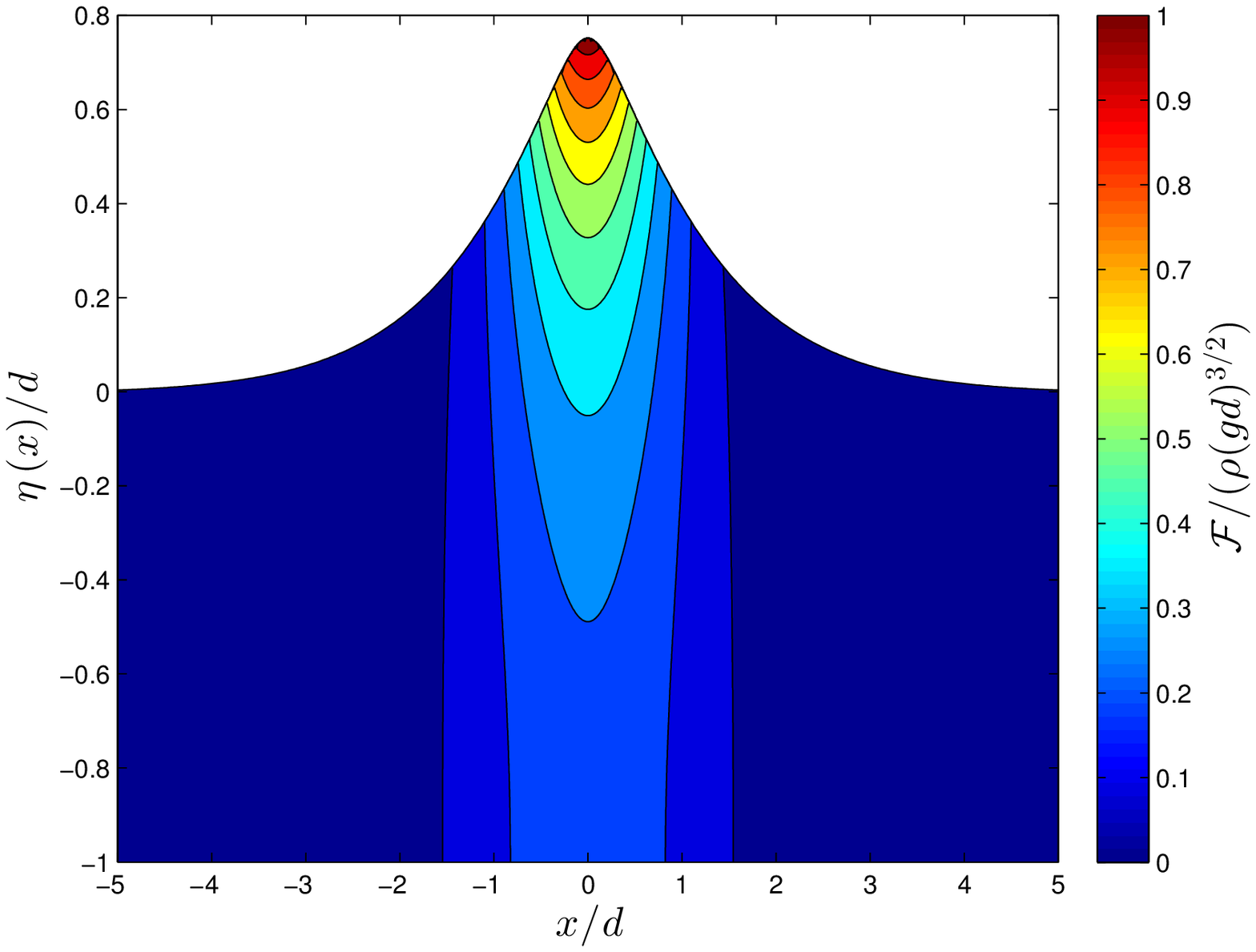}}
  \caption{\small\em Iso-values of the kinetic energy density $\half (u^2+v^2)$ (left) and the total energy flux $\bigl(p + gy + \half(u^2+v^2)\bigr)u$ (right) under a large wave.}
  \label{fig:energy}
\end{figure}

Finally, the Babenko equation allows also to reconstruct efficiently various fields in the bulk of the fluid, as explained in Section~\ref{sec:bulk}. The computation of these fields amounts to perform post-processing operations on the fully converged solution. To illustrate this concept we take the large amplitude solitary wave represented on Figure~\ref{fig:mp}(a). On Figure~\ref{fig:potstream} we show the velocity potential (a) and the stream function (b) inside the fluid. The total (a) and dynamic (b) pressures are shown on Figure~\ref{fig:press}. The horizontal (a) and vertical (b) velocities along with accelerations are represented on Figures~\ref{fig:speeds} and \ref{fig:accel}, correspondingly. Finally, the kinetic energy density (a) along with the total 
energy flux are shown on Figure~\ref{fig:energy}. In particular, one can see from these computations that all presented fields (except the vertical velocity and horizontal acceleration) are symmetric with respect to the wave crest. On the other hand, two remaining quantities are antisymmetric with respect to the vertical axis passing through the crest. These results are in complete agreement with previous theoretical and numerical studies conducted with other methods \cite{Clamond2012a,Constantin2011}.

\section{Conclusions and perspectives}\label{sec:concl}

In the present paper, we proposed a fast and accurate method for computing steady gravity solitary waves to the full Euler equations. The method is based essentially on two main ingredients: first, a conformal mapping in order to reformulate the free surface Euler equations onto a fixed domain; second, the Petviashvili iterations to solve them numerically. The resulting scheme allows to compute the solution to any arbitrary accuracy using the multi-precision floating-point arithmetics \cite{MATLAB2012}. Both ingredients are well known, but their combination turns out to be very efficient and seems to be new. We also compared our solution with some high-order asymptotic expansions \cite{Fenton1972, Longuet-Higgins1974}. We obtain a good agreement for small and moderate amplitude solitary waves. For higher waves, the differences start to be noticeable. Moreover, the proposed method is compared to the classical Tanaka algorithm which is currently widely used in the 
water wave community \cite{Golay2005, CGHHS}. Our method outperforms the Tanaka algorithm in terms of the computational complexity (which results in much shorter CPU times) and the accuracy which is unlimited theoretically, but limited practically by the floating point arithmetics accuracy. The \textsc{Matlab} implementation is rather compact. The computational core is no longer than 50 lines of code. The script is freely available to download and to use for the scientific community through the \emph{Matlab Central File Exchange} server \cite{Clamond2012}.

Concerning the perspectives, the next step will consist in the inclusion of the capillary effects \cite{Dias1999}. This new force introduces some inhomogeneous nonlinearities into the equations. The classical Petviashvili iterations, as it is presented hereinabove, fail to converge for capillary-gravity waves. In a upcoming study, we are going to propose a fix to this problem.

\subsection*{Acknowledgments}
\addcontentsline{toc}{subsection}{Acknowledgments}

D.~\textsc{Dutykh} acknowledges the support from ERC under the research project ERC-2011-AdG 290562-MULTIWAVE. The authors would like to thank also Angel~\textsc{Duran} for very stimulating discussions on the Petviashvili method. Finally, we would like to thank Pavel~\textsc{Holodo\-borodko} for invaluable help with the multi-precision floating point arithmetics.

%%% Bibliography
\addcontentsline{toc}{section}{References}
\bibliographystyle{abbrv}
\bibliography{biblio}

\end{document}